\documentclass[acmsmall]{acmart}

\AtBeginDocument{%
  \providecommand\BibTeX{{%
    \normalfont B\kern-0.5em{\scshape i\kern-0.25em b}\kern-0.8em\TeX}}}

\usepackage[utf8]{inputenc}
\usepackage{graphicx}
\usepackage{subfigure}
\usepackage[ruled,vlined]{algorithm2e}
\usepackage{longtable}
\usepackage[T1]{fontenc}
\usepackage[utf8]{inputenc}
\usepackage{multirow}

\newcommand{\figw}{0.5}
\newcommand{\swap}[2]{\mathrm{SWAP}(#1,#2)}
\newcommand{\cnot}[2]{\mathrm{CNOT}(#1,#2)}

\newcommand{\ket}[1]{\left| {#1} \right\rangle }
\newcommand{\astate}{{\mathbf{s}}}

\newcommand{\num}[1]{\textsc{visit}({#1})}
\newcommand{\scost}[2]{\textsc{scost}(#1,#2)}
\newcommand{\depth}[1]{{\rm I}_{#1}}

\newcommand{\nums}{\textsc{visit}(\astate)}

\newcommand{\vals}{\textsc{val}(\astate)}
\newcommand{\val}[1]{\textsc{val}({#1})}
\newcommand{\rew}[1]{\textsc{rwd}({#1})}

\newcommand{\Nbp}{N_{\textsc{bp}}}
\newcommand{\Nsim}{N_{\textsc{sim}}}
\newcommand{\Gsim}{G_{\textsc{sim}}}
\newcommand{\cexpl}{c}
\newcommand{\discount}{\gamma}
\newcommand{\Lsim}{L_{\textsc{sim}}}

\newcommand{\remote}[2]{\mathcal{R}_{#1}({#2})}
\newcommand{\tket}{t$\ket{\mathrm{ket}}$}

\newcommand{\B}[1]{\mathcal{B}_{#1}}

\usepackage[normalem]{ulem}
\usepackage{xcolor}

\newcommand{\blue}[1]{#1}

\newtheorem{example}{Example}

\SetKwRepeat{Do}{do}{for}

\begin{document}

\title{\blue{Quantum Circuit Transformation: A Monte Carlo Tree Search Framework}}
\author{Xiangzhen Zhou}
\affiliation{%
  \institution{Centre for Quantum Software and Information, Faculty of Engineering and Information Technology, University of Technology Sydney}
  \country{Australia}
  }
 \affiliation{%
  \institution{State Key Lab of Millimeter Waves, Southeast University}
  \country{China}}

\author{Yuan Feng}
\affiliation{%
  \institution{Centre for Quantum Software and Information, Faculty of Engineering and Information Technology, University of Technology Sydney}
  \country{Australia}}
\email{yuan.feng@uts.edu.au}

\author{Sanjiang Li}
\affiliation{%
  \institution{Centre for Quantum Software and Information, Faculty of Engineering and Information Technology, University of Technology Sydney}
  \country{Australia}}
\email{sanjiang.li@uts.edu.au}



\begin{abstract}
In Noisy Intermediate-Scale Quantum (NISQ) era, quantum processing units (QPUs) suffer from, among others, highly limited connectivity between physical qubits. To make a quantum circuit \blue{\emph{effectively}} executable, a circuit transformation process is necessary to transform it\blue{, with overhead cost the smaller the better,} into a functionally equivalent one so that the connectivity constraints imposed by the QPU are satisfied. While several algorithms have been proposed for this goal, the overhead costs are often very high, which degenerates the fidelity of the obtained circuits sharply. One major reason for this lies in that, due to the high branching factor and vast search space, almost all these algorithms only search very \emph{shallowly} and thus, very often, only (at most) locally optimal solutions can be reached. 
In this paper, we propose a Monte Carlo Tree Search (MCTS) framework to tackle the circuit transformation problem, which enables the search process to go much deeper. 
The general framework supports implementations aiming to reduce either the size or depth of the output circuit through introducing SWAP or remote CNOT gates. The algorithms, called MCTS-Size and MCTS-Depth, are polynomial in all relevant parameters. Empirical results on extensive realistic circuits and IBM Q Tokyo show that 
the MCTS-based algorithms 
can reduce the size (depth, resp.) overhead by, on average, 66\% (84\%, resp.) when compared with \tket, an industrial level compiler. 
\end{abstract}

\maketitle

\section{Introduction}
With Google's recent conspicuous, though arguable, success in demonstrating quantum supremacy in a 53-qubit quantum processor \cite{HarrowSupremacy}, NISQ (Noisy Intermediate-Scale Quantum) devices have attracted rapidly increasing interests from researchers in both academic and industrial communities. \emph{Quantum processing units} (QPUs) in the NISQ era only support a limited  set of basic operations (elementary quantum gates) and often suffer from high gate errors, short coherence time, and limited connectivity between physical qubits. In order to run a quantum algorithm, described as a quantum circuit, we need to \emph{compile} the circuit (referred to as \emph{logical circuit} henceforth) into a functionally equivalent \emph{physical circuit} executable on the QPU. 
The compilation includes two basic processes. In the \emph{decomposition} process, gates in the logical circuit are decomposed, or transformed, into elementary gates supported by the QPU \cite{compile,sivarajah2020t,qiskitIBM}. 
The \emph{transformation} process, initiated in \cite{MaslovFM07,cheung2007translation} and also
known as \emph{quantum circuit transformation} (QCT) \cite{Childs} or  \emph{qubit mapping} \cite{LiUCSB}, is then performed on the generated circuit, which further consists of two steps: \emph{initial mapping construction} and \emph{qubit routing}. The former process constructs a mapping that maps qubits in a logical circuit, called \emph{logical qubits}, to the ones in the QPU, called \emph{physical qubits}; while the latter transforms a circuit through adding ancillary operations like SWAP gates to `route' physical qubits in order to make all multi-qubits gates executable.

Both the decomposition and the transformation processes have been studied extensively in the literature. As there are now standard decomposition processes (see,  e.g., \cite[Chapter~4]{Nielsen}), 
in this paper, we focus on the transformation process, 
 and assume that gates in the input logical circuit have been well decomposed into elementary gates that are supported by the QPU. Furthermore, we assume that an initial mapping is given, which can be obtained by employing, say,
 the greedy strategy \cite{Astar,initialmap,CowtanRouting}, the reverse traversal technique \cite{LiUCSB}, the simulated annealing based algorithm \cite{SAHS}, or the subgraph isomorphism based methods \cite{MaslovFM07,siraichi2019qubit,SubgraphIsomorphism}.

To reduce the gate overheads in the qubit routing step, many algorithms have been proposed aiming at minimising gate counts \cite{Astar,SAHS,SubgraphIsomorphism,LyeMinSwap}, circuit depths \cite{laoMapping,USTC,BoothPlanning,VenturelliPlanning} or circuit error \cite{Rodney,qubitmovement}. These algorithms can be roughly classified into two broad categories (see also \cite{Kusyk21-qctsurvey} for a similar classification). 
\blue{The first category consists of algorithms that try to reformulate QCT as a planning or optimisation problem and solve it by applying  off-the-shelf tools \cite{BoothPlanning,VenturelliPlanning,SiraichiAllocation,SaeediLinearNearest,VenturelliPlanner,qubitmovement,AlmeidaPermutations,rasconi2019innovative,zhu2020exact,wille2019mapping,zhang2021time}.} However, as shown in \cite{SiraichiAllocation,Childs},  QCT is NP-complete in general. Algorithms in this category are usually highly unscalable when the size of input circuits becomes large.

In contrast, algorithms in the second category use heuristic search to construct the output quantum circuit step by step from the original input quantum circuit \cite{LiUCSB,initialmap,Astar,SiraichiAllocation,oddi2018greedy,FiniganAllocationNISQ,SAHS}. 
Experimental results show that customised heuristic search algorithms are more promising in transforming large-scale circuits, but usually there is still a considerable gap between the output circuit and an optimal one. The reason partially lies in the limited search depth in most of these algorithms.
To achieve efficiency, one either divides the circuits into layers and tries to execute the gates layer-wise \cite{Astar}, or simply considers only the direct effect of a single move (i.e., SWAP) (see e.g., \cite{LiUCSB,Childs,CowtanRouting}). This leads to a very shallow search depth. The Simulated Annealing and Heuristic Search algorithm (SAHS) \cite{SAHS} and the Filtered and Depth-Limited Search approach (FiDLS) \cite{SubgraphIsomorphism} can go one or two steps further, but exploring even more seems impractical as the searching process will become very slow if many qubit connections are present in the QPU. \blue{Recently, machine learning techniques have also been exploited to provide a more precise evaluation tool for QCT algorithms \cite{QCT_ML,pozzi2020using,MCTS_ML}. Whereas, those algorithms often suffer from a poor scalability in terms of the number of qubits of the NISQ device.}


Inspired by the recent spectacular success of Monte Carlo Tree Search (MCTS) in Computer Go play \cite{alphago1,alphago2}, in this paper, we propose an MCTS framework for the QCT problem. Although first designed for solving computer games, MCTS has found applications in many domains which can be represented as trees of sequential decisions \cite{MCTS2}. 
MCTS is a flexible statistical anytime algorithm, which can be used with little or no domain knowledge \cite{MCTS2}. The basic idea behind MCTS is to explore and exploit, in a balanced way, a search tree in which each node represents a game state and each branch a legal move starting from that state. Given the current game state, the aim is to select the most promising move by exploring a search tree rooted with this state, based on random sampling of the search space.
This is achieved through the following five steps: (1) \emph{Selection}. Starting from the root, we first select successively a child node until a leaf node is reached; (2) \emph{Expansion}. Expand the selected leaf node with one or more child nodes each of which corresponding to a legal move; (3) \emph{Simulation}. Play out the task to completion by selecting subsequent moves randomly; (4) \emph{Backpropagation}. Backpropagate the simulation result (wining, losing, or the reward points collected) towards the root node to update the values of nodes along the way; (5)  \emph{Decision}. After repeated a sufficient number of times, we then select the best move (with the largest value) and move to the next game state.

\begin{figure*}[t]
	\centering
	\includegraphics[width=0.8\textwidth]{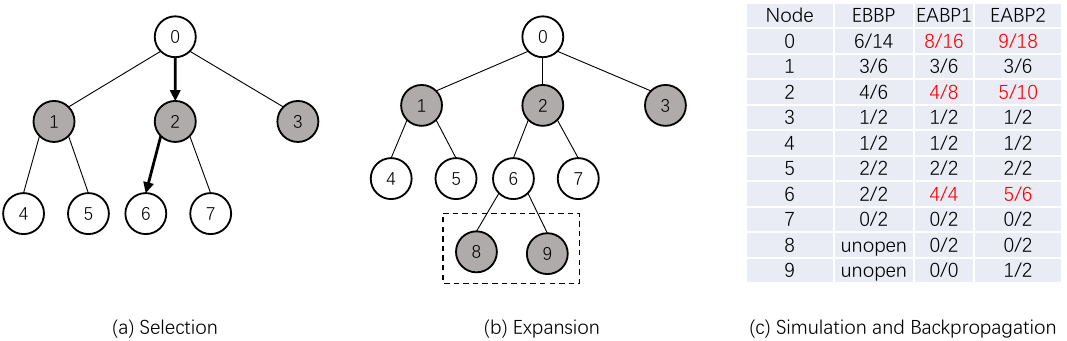}
	\caption{Example for one playout in MCTS. Each grey or white node in (a) and (b) represents a legal move made by the corresponding player. The last three columns in (c) represent, respectively, evaluations in each node before, after the first, and after the second \emph{Backpropagation}. The evaluation is defined as $\#wins / \#simulations$ obtained by \emph{Simulation} and Backpropagation. 
	}
	\label{fig:MCTS_ex1}
\end{figure*}

\begin{example} 
\label{ex:mcts}
We show how to conduct a full playout based on a search tree as shown in Fig.~\ref{fig:MCTS_ex1}(a). Suppose a simple strategy only choosing child with maximum winning rate\footnote{The strategy for \emph{Selection} in practice is much more complex than this and should take both evaluations and time of visits into account. Interested readers can refer to \cite{KocsisUCT,MCTS} for further details.}  is used in \emph{Selection}. Then starting from root node 0, nodes 2 and 6 with maximum {\#wins/\#simulations} values {$4/6$} and $2/2$ among their peers according to the data in column `EBBP', Evaluation Before BackPropagation, in Fig.~\ref{fig:MCTS_ex1}(c) will be chosen successively. Because node 6 is a leaf, it will be expanded and its child nodes 8 and 9, as shown in the dashed box of Fig.~\ref{fig:MCTS_ex1}(b), will be opened. After \emph{Expansion}, one or more newly opened nodes will be chosen to perform simulations. In this example, 
both nodes 8 and 9 are chosen to execute 2 random simulations and the results are assumed to be $0/2$ and $1/2$ respectively. 
After all simulations in node 8 are done, the result will be back propagated to root node 0 through nodes 6 and 2, and their values will be updated and are marked red in the `EABP1', Evaluation After BackPropagation, column of Fig.~\ref{fig:MCTS_ex1}(c). To be specific, the denominator values of nodes 6, 2 and 0 along the backpropagation path will be increased by 2 because the same number of new simulations are done in node 8; only the numerator values of white nodes 6 and 0 are increased because the black player lost both simulations. The same operation applies after the simulations in node 9 are finished and the updated values can be found in the `EABP2' column.
\end{example}

Our MCTS framework for the QCT problem also consists of these five major modules. 
%
%
%
%
In the framework, we adopt a fast random strategy for simulation and carefully design a scoring mechanism which takes both short and long-term rewards into consideration. Based on the five modules and the scoring mechanism, an algorithm, abbreviated as MCTS-Size, is proposed to optimise the size of the output circuit. The algorithm is polynomial in all relevant parameters and experiments on an extensive set of realistic benchmark circuits show that the search depth can easily exceed {most, if not all,} existing algorithms. The search depth in the proposed algorithm is defined as the depth of the selected leaf node to the root during each invoking of the Selection module. In Example~\ref{ex:mcts}, node 6 is chosen in Selection and its search depth is 2. This deep search method can reduce the gate overhead of the output physical circuits by a large margin 
when compared with the state-of-the-art algorithms \cite{SAHS,SubgraphIsomorphism,CowtanRouting} on IBM Q20.

Although aiming to optimise the circuit size in terms of gate numbers, MCTS-Size also reduces the depth of the output circuit significantly.
Depth is perhaps a more significant  criterion for quantum circuits due to the highly limited coherence time in NISQ devices.
When compared with \tket\ introduced in \cite{CowtanRouting}, a state-of-the-art and industrial level algorithm  aiming at depth optimisation, MCTS-Size reduces the circuit size and depth overheads by, respectively, 66\% and 75\% on IBM Q20 (cf. Table~\ref{tab:summary}). More importantly, as our MCTS framework is flexible, it can be easily adapted to accommodate various optimisation criteria. To exemplify this feature, we design MCTS-Depth by introducing two very simple modifications to MCTS-Size. Experimental results on IBM Q20 show that, compared to \tket\ again, MCTS-Depth is able to reduce the depth overhead up to 84\%.

This paper is a significant extension of the conference paper \cite{MCTS_QCT} presented at ICCAD'20. Among others, we have made the following major extensions: (a) aiming to optimise the output circuit depth, we design the MCTS-Depth algorithm (cf. Sec.~\ref{sec:MCTS-Depth}) (note that \cite{MCTS_QCT} only considered  optimisation of the output circuit size); (b) to further demonstrate the flexibility of our framework, we incorporate remote CNOT gates into the MCTS-based algorithms (cf. Sec.~\ref{sec:remote_cnot})\blue{, which are also known as bridge gates and can execute CNOT gates whose two qubits are not neighbours without changing the current mapping}; (c) we describe in detail the parameter selection process and empirically compare the search depth of the MCTS-Size algorithm with that of SAHS \cite{SAHS} (cf. Sec.~\ref{sec:benchmark}); (d) we present detailed and additional empirical evaluation results (considering depth reduction as well as the effect of remote CNOT gates) on \blue{IBM Q20, a hypothetical grid-like QPU called Grid $4\times 5$, and IBM Rochester and Google Sycamore, and with two other state-of-the-art algorithms, viz., Qiskit and SABRE \cite{LiUCSB} (cf. Sec.~\ref{sec:benchmark}).}

The remainder of this paper is organised as follows: Sec.~\ref{sec:background} provides some background knowledge about quantum computation and summarises the state-of-the-art of the quantum circuit transformation problem. Sec.~\ref{sec:MCTS} then presents a detailed description of the MCTS framework as well as a theoretical analysis. The adapted depth-optimisation algorithm is presented in Sec.~\ref{sec:MCTS-Depth}. After that, we show how to incorporate remote CNOT in the MCTS-based algorithms in Sec.~\ref{sec:remote_cnot}. \blue{Empirical evaluations of both MCTS-based algorithms on an extensive set of realistic benchmark circuits and on various QPUs are presented in Sec.~\ref{sec:benchmark}.} The last section concludes the paper with an outlook.

\section{Quantum Circuit Transformation}
\label{sec:background}

In classical computing, data are stored in the form of bits which \blue{can take one of} two states, 0 and 1. In contrast, data  in quantum computing are stored in \emph{qubits}, which also have two basis states represented by $\ket{0}$ and $\ket{1}$, respectively. However, unlike a classical bit, a qubit can be in the superposition
$\alpha \ket{0} + \beta \ket{1}$ 
of basis states,
where $\alpha$ and $\beta$ are complex numbers satisfying ${\left| \alpha  \right|^2} + {\left| \beta  \right|^2} = 1$.

\begin{figure}[t]
	\centering
	\includegraphics[width=0.3\textwidth]{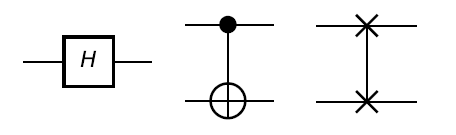}
	\caption{Hadamard, CNOT and SWAP gates (from left to right).}
	\label{fig:gates}
\end{figure}

\begin{figure}[t]
	\centering
	\includegraphics[width=0.4\textwidth]{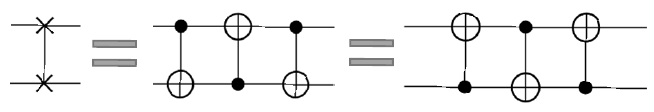}
	\caption{The decomposition of a SWAP into three CNOT gates.}
	\label{fig:gate_de}
\end{figure}

The state of a qubit can be changed by quantum gates, which are mathematically represented by unitary matrices. Fig.~\ref{fig:gates} depicts three important quantum gates used in this paper: Hadamard, CNOT and SWAP gates. Hadamard is a single-qubit gate that has the ability to generate superposition: it maps $\ket{0}$ to $(\ket{0}+\ket{1})/\sqrt{2}$ and $\ket{1}$ to $(\ket{0}-\ket{1})/\sqrt{2}$. CNOT and SWAP are both two-qubit gates. CNOT flips the target qubit depending on the state of the control qubit; that is, CNOT: $\ket{c}\ket{t} \rightarrow \ket{c}\ket{c\oplus t}$, where $c,t\in \{0,1\}$ and $\oplus$ denotes exclusive-or. SWAP exchanges the states of its operand qubits: it maps $\ket{a}\ket{b}$ to $\ket{b}\ket{a}$ for all $a,b\in \{0,1\}$. Note that a SWAP gate can be decomposed into three CNOT gates as shown in Fig.~\ref{fig:gate_de}.

\begin{figure*}[t]
	\centering
	\includegraphics[width=0.7\textwidth]{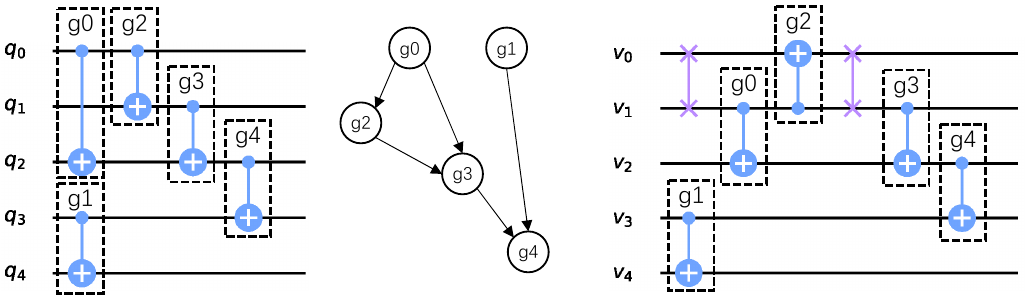}
	\caption{{A quantum circuit (left), its dependency graph (middle) and the circuit after transformation (right).}}
	\label{fig:fig_cir_DG}
\end{figure*}

Quantum gates can be concatenated to form complex \emph{circuits} which, together with measurements, are used to describe quantum algorithms. A circuit is usually denoted by a pair $\left( {Q,C} \right)$, where $Q$ is a set of qubits and $C$ a sequence of quantum gates on $Q$. Sometimes we also call $C$ a circuit when $Q$ is clear from the context. 
Fig.~\ref{fig:fig_cir_DG} shows a circuit where
$Q=\left\{ {{q_0}, \ldots, {q_4}} \right\}$, $C = \left( {{g_0}, \ldots ,{g_4}} \right)$, $g_0 = \cnot{q_0}{q_2}$, $g_1 = \cnot{q_3}{q_4}$, etc. Here each CNOT is annotated with the qubits on which they are applied.

\subsection{Quantum Circuit Transformation\blue{: Problem Formulation} }
\label{sec:QCT}
As mentioned in the introduction, to run a quantum circuit on a given QPU in the NISQ era,
we need to transform it so that the connectivity constraints imposed by the QPU are all satisfied. Such connectivity constraints are typically described as
an undirected and connected graph $AG = \left( {V,E} \right)$, {called} the \emph{architecture graph} \cite{Childs}, where $V$ denotes the set of physical qubits of the QPU and $E$ the pairs of physical qubits on which a two-qubit gate can be applied.

\begin{figure}[t]
	\centering
	\includegraphics[width=0.75\textwidth]{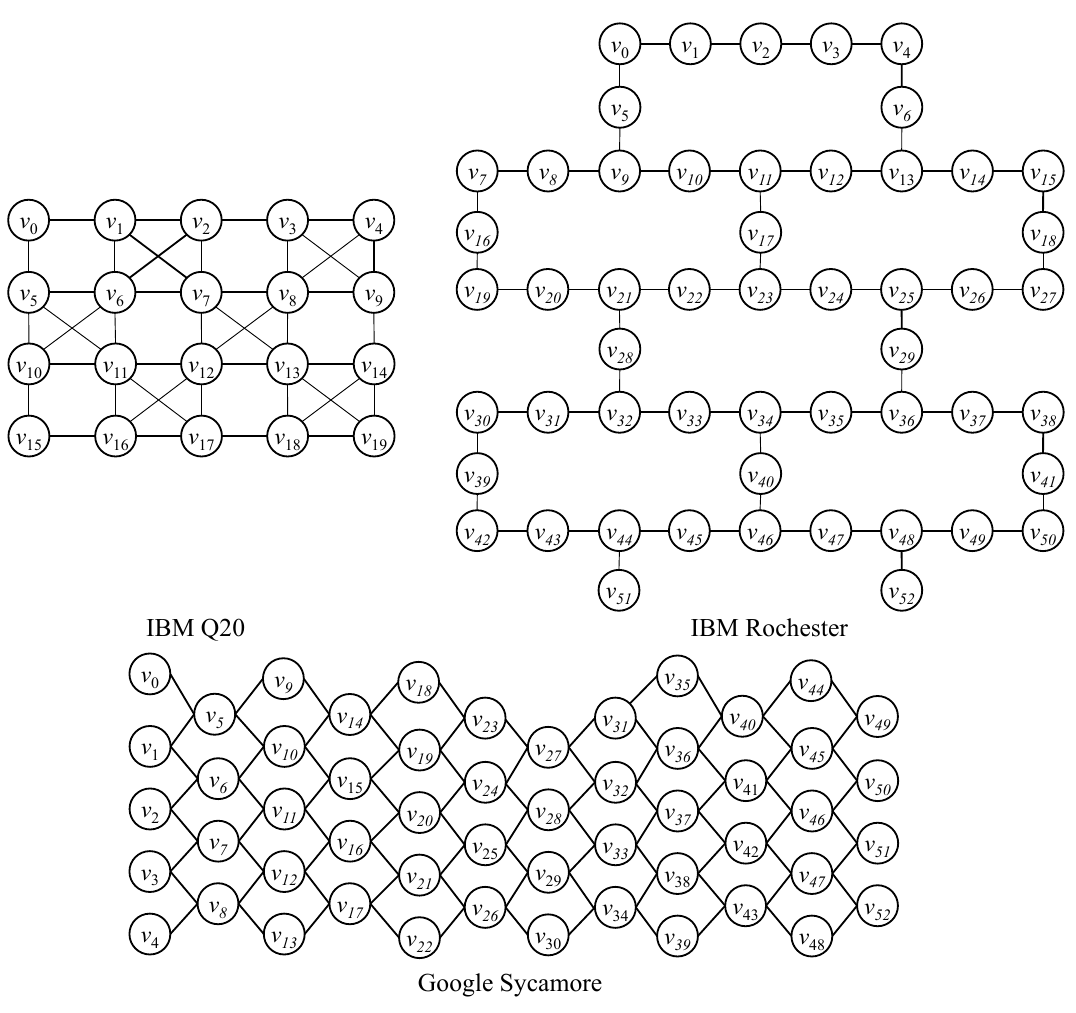}
	\caption{\blue{The architecture graphs for IBM Q20, IBM Rochester and Google Sycamore.} 
	}
	\label{fig:AG}
\end{figure}





	
Note that by a standard process \cite{Nielsen}, any quantum circuit can be decomposed into a functionally equivalent one which consists of only CNOT and single-qubit gates. Furthermore, as single-qubit gates can be executed directly on a QPU (connectivity constraints only prevent two-qubit gates from applying on certain pairs of physical qubits), 
if not otherwise stated, we assume that single-qubit gates have been removed and the circuit to be transformed consists solely of CNOT gates.\footnote{This implies that we cannot simplify the circuits by, say, cancelling two consecutive CNOT gates acting on the same pair of qubits.}  
Note that this is only a technical assumption and, whenever necessary, we can always add back the corresponding single-qubit gates (cf. Sec.~\ref{sec:MCTS-Depth}). Besides that, the SWAP gates added during the transformation process will be decomposed into CNOTs in the end.
	


An important notion related to quantum circuits which plays a key role in QCT is the \emph{dependency graph}.
Let $C = \left( {{g_0},{g_1}, \ldots } \right)$ be a quantum circuit.
We say gate $g_i$ in $C$ \emph{depends} on $g_j$ if $j < i$ and they share at least one common qubit. The dependence is  \emph{direct} if there is no gate $g_k$ with $j < k < i$ such that $g_i$ depends on $g_k$ and $g_k$ depends on $g_j$.
In general, we can construct a directed acyclic graph (DAG), called the {dependency graph} \cite{DG}, to characterise the dependency between gates in a circuit. Specifically, each node of the dependency graph represents a gate and each directed edge the {direct} dependency relationship between the gates involved.
Let's say a gate $g_i$ directly depends on $g_j$, then the corresponding edge $(g_j, g_i)$ should be added to the dependency graph.
With the help of dependency graph, any quantum circuit $C$ can be divided into different \emph{layers} {such that gates in the same layer can be executed in parallel}. The first or front layer, denoted by $\mathcal{L}_0 (C)$, consists of the gates which have no parents in the DAG. The second layer, $\mathcal{L}_1 (C)$, is then the front layer of the DAG obtained by deleting all gates in $\mathcal{L}_0 (C)$. Analogously, we can define the $i$-th layer of a circuit for any $i \ge 0$.

\begin{example}\label{ex:qct-remote-cnot}
{Fig.~\ref{fig:fig_cir_DG} shows an example of a quantum circuit (left) and its  dependency graph (right), from which we can see that the front layer of the circuit consists of $g_0$ and $g_1$, the second $g_2$, the third $g_3$, and the fourth $g_4$.}
\end{example}

Another key notion for QCT is a qubit mapping $\tau$ which allocates logical qubits $Q$ to physical qubits $V$ so that for any $q_i, q_j \in Q$, $\tau(q_i) = \tau(q_j)$ if and only if $i=j$. Given a logical circuit $(Q, C)$ and an architecture graph $AG$,
a two-qubit gate $g=\cnot{q_i}{q_j}$ in $C$ is called \emph{executable} by $\tau$ if $\tau(q_i)$ and $\tau(q_j)$ are adjacent in $AG$, and $g$ is either in the front layer of $C$ or all the gates it depends on are executable. Note that in general it is impossible that all two-qubit gates in a circuit are executable by a single mapping. Once no gates are executable by the current mapping $\tau$, a QCT algorithm seeks to insert into the circuit some ancillary SWAP gates to change $\tau$ into a new one so that more gates are executable. This insertion-execution process is iterated until all gates from the input circuits are executed. To illustrate the basic ideas, we revisit the circuit on the left side of Fig.~\ref{fig:fig_cir_DG}.


\begin{example}\label{exa:qct}
	We transform the logical circuit $LC=(Q, C^l)$ shown in Fig.~\ref{fig:fig_cir_DG} into a physical one $PC=(V, C^p)$ satisfying the architecture graph $AG$ in Fig.~\ref{fig:AG}. Suppose the initial qubit mapping $\tau$ is given as a naive one which maps $q_i$ to $v_i$, $0\leq i\leq 4$. 
	\begin{enumerate}
		\item Since $\tau(q_3) = v_3$, $\tau(q_4) = v_4$, and $v_3$ and $v_4$ are adjacent in $AG$, gate $g_1$ in $C^l$ is already executable by $\tau$. Thus we initialise $PC$ as a physical circuit with $V=\{v_0, \ldots, v_{19}\}$ containing only a single CNOT gate acting on $v_3$ and $v_4$, and delete $g_1$ from $C^l$. Thus now, $C^l = (g_0, g_2, g_3, g_4)$ and $C^p = (\cnot{v_3}{v_4})$.
		\item As no gates in $C^l$ is executable by $\tau$, we have to insert a SWAP (or a sequence of them) to get a new mapping which admits more CNOT gates from $C^l$ executable. In this example, we choose to add $\swap{v_0}{v_1}$ to $C^p$, which in effect converts $\tau$ into $\tau'$ that maps $q_0$ to $v_1$ and $q_1$ to $v_0$.
		Now $g_0$, which acts on $q_0$ and $q_2$, is executable (since $v_1$ and $v_2$ are adjacent in AG). Similarly, $g_2$ is executable as well. Thus they can be deleted from $C^l$ and added into $C^p$ (with the operand qubits changed accordingly). Consequently, now $C^l = (g_3, g_4)$ and
		\begin{align*}
		C^p = ( \cnot{v_3}{v_4}, \swap{v_0}{v_1}, \cnot{v_1}{v_2},\cnot{v_1}{v_0} ).
		\end{align*} 
		\item Proceeding in a similar way, we add another $\swap{v_0}{v_1}$ to $C^p$ to converts $\tau'$ back to $\tau$ so that $g_3$ and $g_4$ are executable. After deleting them from $C^l$ and adding them into $C^p$, we have $C^l = \emptyset$ and the finial physical circuit becomes
		\begin{align*}
			C^p &=  \big(\cnot{v_3}{v_4}, \swap{v_0}{v_1}, \cnot{v_1}{v_2}, \cnot{v_1}{v_0},\\
			&\quad\quad\quad\quad \swap{v_0}{v_1}, \cnot{v_1}{v_2}, 
			\cnot{v_2}{v_3}\big),
		\end{align*} 
		which satisfies all the connectivity constraints of AG. {The final physical circuit is shown in Fig.~\ref{fig:fig_cir_DG} (right).}
	\end{enumerate}
%
%
%
\end{example}

%

%

\subsection{Heuristic Search Algorithms}



Recall that given a logical circuit $LC_0$, an architecture graph $AG$, and an initial qubit mapping $\tau_{ini}$, the QCT process aims to output a physical circuit which respects all the connectivity constraints in $AG$.
To present this process as a search problem, we need to first define the notion of states. Naturally, a \emph{state} of the QCT process is a triple $\astate=(\tau, PC, LC)$, where $\tau$ is a qubit mapping describing the current allocation of logical qubits, $PC$ is the physical circuit that consists of all gates that have been executed so far and the auxiliary SWAP gates inserted and the logical circuit $LC$ consists of the remaining gates to be executed. Sometimes we denote by $LC(\astate)$ and $PC(\astate)$ the logical and the physical circuits of $\astate$, respectively.

A \emph{legal action} in the QCT process can be either a SWAP operation (corresponding to an edge in $AG$) or a sequence of SWAP operations.\footnote{In Sec.~\ref{sec:remote_cnot} we will relax this restriction and allow remote CNOTs to be legal actions.}  Let $\astate=(\tau, PC, LC)$ be the current state, and suppose an action $\swap{v_i}{v_j}$ is taken on $\astate$. Then a new state $\astate'=(\tau', PC', LC')$ is reached where $\tau'$ is the same as $\tau$ except that it maps $\tau^{-1}(v_i)$ to $v_j$ and $\tau^{-1}(v_j)$ to $v_i${, where $\tau^{-1}(v_i)$ and $\tau^{-1}(v_j)$ are, respectively, the preimages of $v_i$ and $v_j$ under $\tau$}. Furthermore, $LC'$ is obtained from $LC$ by deleting all gates which are executable by $\tau'$, and $PC'$ is obtained from $PC$ by adding first $\swap{v_i}{v_j}$ and then all the gates just deleted from $LC$, with the operand qubits changed according to $\tau'$. While most algorithms select one SWAP each time, the $A^*$ algorithm \cite{Astar} and FiDLS \cite{SubgraphIsomorphism} select a sequence of SWAPs. {Note that when regarding sequences of SWAPs as legal actions, usually we execute a gate only after the last SWAP is applied.} 

The \emph{initial state} $\astate_0$ of the QCT process is taken as $(\tau_{ini}, PC_0, LC_0')$ where $PC_0$ is the physical circuit consisting of all gates from $LC_0$ which are executable by $\tau_{ini}$, and $LC_0'$ the logic circuit obtained by deleting all gates in $PC_0$ from $LC_0$. The \emph{goal states} are those with the associated logical circuit being empty. Note that the associated physical circuit of any goal state respects the connectivity restraints in $AG$.
The \emph{cost} of a state $\astate{}$ depends on the optimisation objective. In this paper, it can be either the total number of auxiliary gates inserted or the depth overhead of the stored physical circuit of $\astate{}$. The aim of QCT is to find a goal state with the minimal cost w.r.t. the particular objective.

Many QCT algorithms in the literature adopt a divide-and-conquer approach in the search process. Starting from the current state $\astate = (\tau, PC, LC)$, each subtask consists of executing the front layer, the first two layers, or a front section of the circuit. For example, in the $A^*$ algorithm, a shortest path in $AG$ (which corresponds to a sequence of SWAPs) is found which converts $\tau$ to a new mapping so that all gates in the first two layers of $LC$ are executable. In \cite{CowtanRouting}, Cowtan et al. partition $LC$ into layers and then
select the SWAP which can maximally reduce the diameter of the subgraph composed of all pairs of qubits in the current layer. Siraichi et al. \cite{SiraichiSCP19} decompose  $LC$  into sub-circuits each of which leads to an isomorphic subgraph of $AG$ and thus the corresponding embedding can act as a mapping $\tau'$ that executes all gates in the sub-circuit. Their algorithm then tries to find a minimal sequence of SWAPs which converts $\tau$ to $\tau'$. A similar approach is also adopted in Childs et al. \cite{Childs}.

Unlike the above algorithms, SAHS \cite{SAHS} and FiDLS \cite{SubgraphIsomorphism} do not divide the problem into sub-problems. Whenever a mapping is generated, they try to execute as many as possible gates from the logical circuit, no matter which level they are in. SAHS regards each SWAP as a valid action, but when selecting the best SWAP to enforce, it simulates the search process one step further and select the SWAP which has the best consecutive SWAP to apply. {In principle, SAHS can go deeper but this will make the algorithm much slower (cf. Fig.~\ref{fig:depth_SAHS}} for an example). FiDLS regards any sequence with up to $k$ SWAPs as a legal action and selects the sequence which executes the most number of gates per SWAP. In a sense, this means that its search depth can reach $k$. To ensure the running time is acceptable, in the experiments on Q20, FiDLS chooses $k$ as 3 and introduces various filters to filter out unlike SWAPs.  

\section{The Proposed MCTS Framework}
\label{sec:MCTS}

In this section, we describe an MCTS framework for quantum circuit transformation and present a detailed algorithm implementation. The algorithm, called MCTS-Size, aims at finding a goal state which has the minimal number of SWAPs inserted. Shortly in Sec.~\ref{sec:MCTS-Depth} we shall see this can be easily adapted to address other optimisation objectives.

Like general MCTS algorithms, our framework also consists of five major parts: \emph{Selection}, \emph{Expansion}, \emph{Simulation}, \emph{Backpropagation} and \emph{Decision}. However, some significant modifications have been made to cater to the unique characteristics of QCT.

\begin{figure*}[t]
	\centering
	\includegraphics[width=0.9\textwidth]{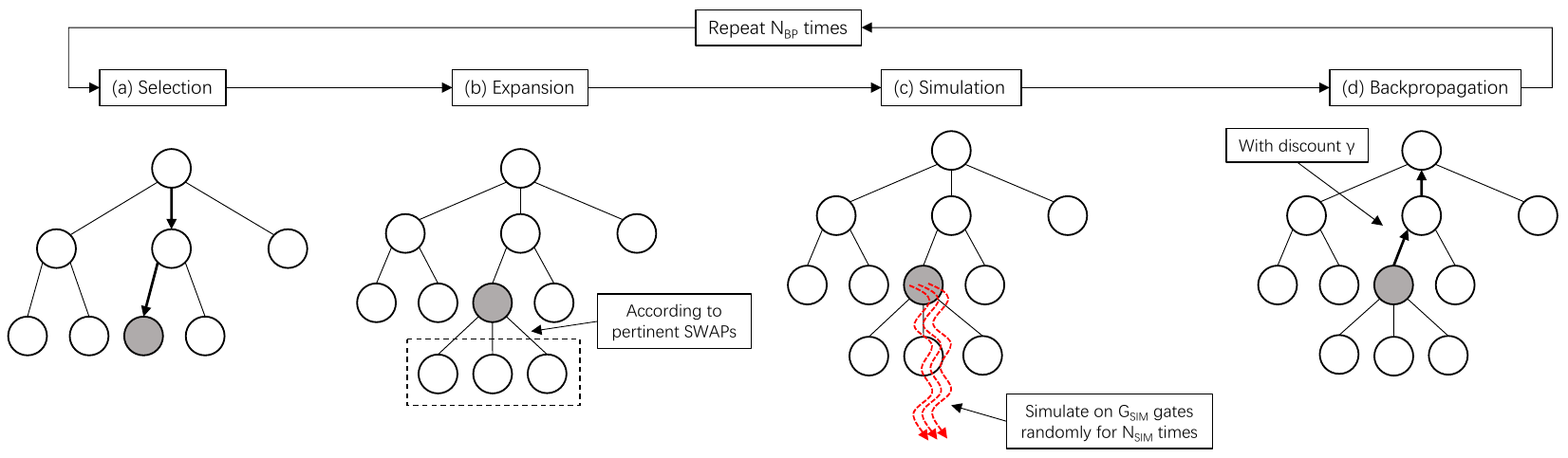}
	\caption{Overview of the Monte Carlo Tree Search Framework. }
	\label{fig:overview}
\end{figure*}

The Monte Carlo search tree for QCT, which is initialised immediately after the algorithm starts, stores all states having been explored during the transformation process. \blue{In practical implementation, it is not necessary to store the full physical and logical circuits in a state; instead, only the incremental information, i.e., the gates added to and removed from, respectively, the physical and logical circuits of its parent state, are stored. When necessary, the circuits of a state can be restored from the incremental information in itself and its ancestor states.}
As stated in the previous section, an edge $\left(\astate', \astate \right)$ connecting node $\astate'$ and its child $\astate$ indicates that a SWAP is applied to convert $\astate'$ to $\astate$. With the aim to minimise the number of inserted gates, we define an immediate, short-term \emph{reward} for each edge and a long-term \emph{value} for each node of the search tree as follows.

\textbf{The short-term reward} $\rew{\astate', \astate}$ is the reward collected from the parent node $\astate'$ to the child $\astate$, in terms of the number of gates executed by the newly inserted SWAP when this transition is made: 
\begin{equation}\label{eq:local_score}
\rew{\astate', \astate} = \mbox{\#gates$\_$in$\_$ $LC(\astate')$  - \#gates$\_$in$\_$$LC(\astate)$}.
\end{equation} 

\textbf{The long-term value} $\vals$. To determine the value of a state $\astate$, the following two factors are taken into account: (i) the (inverse of the) number of inserted SWAPs when transformation of the remaining logical circuit is simulated at $\astate$.
For efficiency, the simulation is performed on,  instead of $LC(\astate)$ itself, a fixed-size sub-circuit of $LC(\astate)$. It is expected that the larger the sub-circuit is for simulation, the better simulated value will be obtained. (ii) the (simulated) value of its best child node and the reward to it collected from $\astate$. To be specific, 
\begin{equation*} 
\vals = \max \{\textsc{sim}, \discount \cdot [\rew{\astate, \astate''}  + \val{\astate''}]\},
\end{equation*}
 where $\textsc{sim}$ is the simulated value obtained from (i), $\astate''$ is the child of $\astate$ with the maximal value, and $\discount$ is a predefined discount factor satisfying $\discount < 1$. In our later implementation, $\vals$ is initially assigned $\textsc{sim}$ in the \emph{Simulation} module, and then updated in \emph{Backpropagation}, whenever simulations are performed at a descendant of $\astate$. Intuitively, $\vals$ describes the efficiency of introducing SWAPs (in terms of the average number of executed gates per SWAP) from $\astate$, considering both the simulation at itself and the backpropagated one from this child nodes. Obviously, the larger $\vals$ is, the smaller the number of SWAPs needed to lead $\astate$ to a goal node, and the `better' $\astate$ is (compared with its siblings).

In addition to the above definitions, as shown in Fig.~\ref{fig:overview}, our framework differs from traditional MCTS algorithms for game playing in the following ways:
\begin{enumerate}
	\item The simulation is performed on the leaf node selected in the \emph{Selection} module, instead of the child nodes opened in the \emph{Expansion} one. Experimental results on real benchmarks indicate that this achieves a better performance for the QCT problem. 
	\item In game playing, the simulation result can be obtained only when the game is decided. In contrast, the reward of a move in our setting is collected during the execution of CNOT gates from the logic circuit. Consequently, in the \emph{Simulation} module, we simulate only on a sub-circuit of the current logic circuit to improve efficiency.
	\item We introduce a discount factor, which can be adjusted to better suit the problem setting, when backpropagating the simulated values.
\end{enumerate}

\subsection{Main modules}\label{sec:modules}
We now elaborate the five major modules one by one.

\vspace{1em}
\textbf{Selection. }\emph{Selection} is the iterated process to find an appropriate leaf node in the search tree to expand and simulate. 
It starts from the root node and, in each iteration, evaluates and picks one of the child nodes until a leaf node is reached.

The way we evaluate child nodes during \emph{Selection} is critical to the performance of the whole algorithm. On one hand, if we only consider their values, the chance for exploring unpromising nodes will be too low and we can easily get stuck in a local minimum. On the other hand, if we always select nodes with a smaller visit count, the search will be too shallow and thus a large amount of time will be wasted in exploring inferior nodes. To get a balance between these two aspects, the following evaluation formula, similar to the well-known  UCT (Upper Confidence Bound 1 applied to trees)  \cite{KocsisUCT}, is introduced in our implementation to make a balanced evaluation among all child nodes $\astate'$ of $\astate$:
\begin{equation}
\label{eq:selection}
    \rew{\astate, \astate'} + \val{\astate'} +  {\cexpl}\sqrt{\frac{\log \textsc{visit}(\astate)} {\num{\astate'}}}
\end{equation}
where $\cexpl$ is a pre-defined parameter, and  $\nums$ is the number of times that $\astate$ has been visited. 
Intuitively, the first two terms in Eq.~\eqref{eq:selection} correspond to the exploitation rate and the third the exploration rate in UCT. In each iteration of the \emph{Selection} module, the node which maximises Eq.~\eqref{eq:selection} is selected. The \emph{Selection} module is presented in Alg.~\ref{alg:selection}.

\begin{algorithm}[ht]
	\SetKwData{Left}{left}\SetKwData{This}{this}\SetKwData{Up}{up}
	\SetKwFunction{Union}{Union}\SetKwFunction{FindCompress}{FindCompress}
	\SetKwInOut{Input}{input}\SetKwInOut{Output}{output}
\hspace*{-4mm}		\Input{A Monte Carlo search tree $\mathcal{T}$.}
\hspace*{-4mm}		\Output{{A leaf node to expand and simulate}.}
	\caption{Select($\mathcal{T}$)}
	\label{alg:selection}
		
		$\astate \leftarrow root(\mathcal{T})$\;
		$\nums \leftarrow \nums + 1$\;
		\While{\text{$\astate$ is not a leaf node}}
		{
			$\astate \leftarrow \text{{the child node $\astate'$ of $\astate$} that maximises Eq.~\eqref{eq:selection}}$\;
			$\nums \leftarrow \nums + 1$\;
		}
		\Return{$\astate$};
\end{algorithm}

\textbf{Expansion. }The goal of \emph{Expansion} is to open all child nodes of a given leaf node {by applying all relevant SWAP operations}. Given a logic circuit $C$ and a qubit mapping $\tau$,
the set of \emph{pertinent} SWAPs, denoted $\mathcal{SWAP}_{C, \tau}$, is the set of gates $\swap{v_i}{v_j}$ such that either $\tau^{-1}(v_i)$ or $\tau^{-1}(v_j)$ appears in a gate in the current front layer of $C$, i.e.,
\begin{equation*}
    \mbox{$(v_i, v_j) \in E$ and ($\tau^{-1}(v_i) \in Q_0$ or $\tau^{-1}(v_j) \in Q_0$)},
\end{equation*}
{where $Q_0$ is the set of logical qubits that are involved in the gates in $\mathcal{L}_0(C)$. To expand a selected node $\astate=(\tau, PC, LC)$, only gates in $\mathcal{SWAP}_{LC, \tau}$ will be applied to generate child nodes. This strategy has been widely used in quantum circuit transformation, see, e.g., \cite{Astar,LiUCSB,SAHS}. In particular, several variants are introduced in FiDLS \cite{SubgraphIsomorphism}.}

For each pertinent SWAP of $\astate$, a new child node $\astate'$ will be generated. Furthermore, the reward $\rew{\astate,\astate'}$ is as defined in Eq.~\eqref{eq:local_score} and both $\val{\astate'}$ and $\num{\astate'}$ are set as 0. The details can be found in Alg.~\ref{alg:expansion}.

\begin{algorithm}[ht]
	\SetKwData{Left}{left}\SetKwData{This}{this}\SetKwData{Up}{up}
	\SetKwFunction{Union}{Union}\SetKwFunction{FindCompress}{FindCompress}
	\SetKwInOut{Input}{input}\SetKwInOut{Output}{output}
\hspace*{-4mm}		\Input{A Monte Carlo search tree $\mathcal{T}$ and node $\astate = (\tau, PC, LC)$.}
	\caption{Expand($\mathcal{T}, \astate$)}
	\label{alg:expansion}
		\For{all {$\swap{v_i}{v_j}$} in $\mathcal{SWAP}_{LC, \tau}$}
		{
			$\tau' \leftarrow \tau[\tau^{-1}(v_i)\mapsto v_j, \tau^{-1}(v_j)\mapsto v_i]$\;
			$C \leftarrow$ the set of all $\tau'$-executable gates in $LC$\;
			$LC' \leftarrow LC$ with all gates in $C$ deleted\;
			$PC' \leftarrow PC$ by adding $\swap{u_i}{u_j}$ and all gates in $C$\;
			$\astate' \leftarrow (\tau', PC', LC')$\;
			$\val{\astate'},\ \num{\astate'} \leftarrow 0$\;
			Add $\astate'$ as a child node of $\astate$\;
			$\rew{\astate,\astate'} \leftarrow \mbox{number of gates in } C$\;
		}
\end{algorithm}

\vspace{1em}
\textbf{Simulation. }The objective here is to obtain a {simulated score}, serving as the initial long-term value $\vals$, of the current state $\astate$ by simulation.
In our implementation, we perform simulation on the first $\Gsim$, a predefined number, gates in the current logical circuit. While almost all existing QCT algorithms can be {used for this purpose}, for the sake of efficiency, a fast random simulation is designed in Alg.~\ref{alg:simulation}. \blue{Related to this,  in Sec.~\ref{sec:exp_more_ag}, we shall see an MCTS algorithm with a deterministic simulation module.}

Given the current state $\astate$, let $N$ be, among all $\Nsim$ (a predefined number) iterations,  the minimal number of SWAP gates we have inserted until all the first $\Gsim$ CNOT gates of $LC(\astate)$ have been executed. Then the initial long-term value of $\astate$ is defined as  
\begin{equation}
\label{eq:simu_score}
\vals = \discount^{N/2} \cdot \Gsim,
\end{equation}
{where $\discount < 1$} is a predefined discount factor. \blue{What deserves explanation is the way we compute the simulated score (or, the initial value) for state $\astate$ in Eq.~\eqref{eq:simu_score}. 
In particular, one may wonder why we take as the exponent $N/2$ instead of $N$? The intuitive meaning of this definition is as follows. Although these $\Gsim$ gates are executed in different steps during the simulation, for simplicity, we suppose they are {all} executed right at the middle point $\astate'$ which is the $N/2$-generation child of $\astate$. Then the reward collected at the transition to $\astate'$ from its parent is {exactly} $\Gsim$. Note that every edge along the path from $\astate$ to the parent of $\astate'$ has zero reward. Thus, we need only backpropagate the reward collected at $\astate'$ upwards with discount factor $\discount$. This gives the simulated score $\discount^{N/2} \cdot \Gsim$ for $\astate$ as specified in Eq.~\eqref{eq:simu_score}. Real benchmark experiments also confirm that the current choice performs better than simply letting $\vals$ be the sum of all the (discounted) rewards collected during the actual execution of these $\Gsim$ CNOT gates.}


{We next show how to do random simulation.  
Let $C$ be a sub-circuit of $LC(\astate)$ and $\tau$ the current mapping. We write   $\mathcal{SWAP}_{C,\tau}$ for} the set of pertinent SWAPs for $C$ under $\tau$. For any $h \in \mathcal{SWAP}_{C,\tau}$, its \emph{impact factor} is defined as
\begin{equation}
\label{eq:sim_if}
\mathrm{IF}(h): = 
f\Big(
\sum\limits_{g \in \mathcal{L}_0(C)} \scost{g}{\tau} -  
\sum\limits_{g \in \mathcal{L}_0(C)} \scost{g}{\tau'}
\Big)
\end{equation}
\blue{where $\tau'$ is the mapping obtained from $\tau$ after applying $h$;  $\scost{g}{\delta}$ for $\delta=\tau$ or $\delta=\tau'$ is the \emph{swap cost} of $g=\cnot{q_j}{q_k}$ with respect to mapping $\delta$, defined as the shortest distance between the physical qubits $\delta(q_j )$ and $\delta(q_k)$ in the architecture graph, in which the edges have an uniform weight of 1 and thus the distance is the summed edge weights; and $f$ the scaling function defined as}
\begin{equation*}
f \left( x \right) =
\left\{ {\begin{array}{*{20}{cl}}
	0, & \text{ if }x < 0\\
	0.001, & \text{ if }x = 0\\
	x, & \text{ if } x > 0
	\end{array}} \right.
\end{equation*}
\blue{which is slightly different from the Relu function in that it returns a tiny positive value (0.001 in our case) instead of 0 when $x=0$. We make this change to ensure that SWAPs which increase the cost will not be selected when no SWAP can decrease the cost.}


{Then, a probability distribution is obtained as follows} 
\begin{equation}
\label{eq:simulation}
P\left({X = h} \right) = \frac{\mathrm{IF}(h)}{\sum \{\mathrm{IF}(h') \mid h' \in \mathcal{SWAP}_{C,\tau}\} },
\end{equation}
through which a SWAP operation can be sampled from $\mathcal{SWAP}_{C,\tau}$ and used to execute gates from $LC$. Note that this simulation process will be repeated for $\Nsim$, also a predefined parameter, times to obtain the best score.

%

\begin{algorithm}[ht]
	\SetKwData{Left}{left}\SetKwData{This}{this}\SetKwData{Up}{up}
	\SetKwFunction{Union}{Union}\SetKwFunction{FindCompress}{FindCompress}
	\SetKwInOut{Input}{input}\SetKwInOut{Output}{output}
\hspace*{-4mm}	\Input{A Monte Carlo search tree $\mathcal{T}$ and node $\astate=(\tau, PC, LC)$.}
	\caption{Simulate($\mathcal{T}, \astate$)}
	\label{alg:simulation}
		$N \leftarrow \infty$\;
		\Do{$\Nsim$ times}
		{
			$C \leftarrow$ circuit with the first $\Gsim$ gates in $LC$\;
			$n \leftarrow 0$;			$\tau' \leftarrow \tau$\;
			\While{$C$ is not empty}
			{
				Sample $h$ from $\mathcal{SWAP}_{C,\tau'}$ according to the probability distribution in Eq.~\eqref{eq:simulation}\;
				$\tau' \leftarrow \tau'$ by applying $h$\;
				$C \leftarrow C$ with all $\tau'$-executable gates  deleted\;
				$n \leftarrow n+ 1$\;
			}
			\If{$n < N$}
			{$N\leftarrow n$\;}
		}
		$\vals \leftarrow \discount^{N/2} \cdot \Gsim$;
\end{algorithm}


\vspace{1em}

\textbf{Backpropagation. } The \emph{Backpropagation} module updates the values of ancestors of the just simulated node in the search tree.
More precisely, the value of node $\astate$ in the propagated path will be updated as
\begin{equation}
\label{eq:BP}
    \vals  \leftarrow \max\big\{\vals,\ \discount \cdot[\rew{\astate, \astate'} + \val{\astate'}]\big\},
\end{equation}
in which $\astate'$ is the child node of $\astate$ on the path. This reflects the intuitive meaning of $\vals$ discussed at the beginning of this section. The implementation is shown in Alg.~\ref{alg:BP}.

\begin{algorithm}[ht]
	\SetKwData{Left}{left}\SetKwData{This}{this}\SetKwData{Up}{up}
	\SetKwFunction{Union}{Union}\SetKwFunction{FindCompress}{FindCompress}
	\SetKwInOut{Input}{input}\SetKwInOut{Output}{output}
\hspace*{-4mm}	\Input{A Monte Carlo search tree $\mathcal{T}$ and node $\astate$.}
	\caption{Backpropagate($\mathcal{T}, \astate$)}
	\label{alg:BP}
	\While{$\astate \ne root(\mathcal{T})$ }
	{
	    $\astate' \leftarrow$ parent node of $\astate$\;
	    $\val{\astate'} \leftarrow \max \{\val{\astate'}, \discount \cdot [\rew{\astate', \astate} + \vals]\}$\;
	   $\astate \leftarrow\astate'$\;
	}
\end{algorithm}

\textbf{Decision.} This module, depicted in Alg.~\ref{alg:decision}, decides the best move from the root node $rt$ and updates the search tree with the subtree rooted at 
the best child node of $rt$.


\begin{algorithm}[ht]
	\SetKwData{Left}{left}\SetKwData{This}{this}\SetKwData{Up}{up}
	\SetKwFunction{Union}{Union}\SetKwFunction{FindCompress}{FindCompress}
	\SetKwInOut{Input}{input}\SetKwInOut{Output}{output}
	\hspace*{-4mm}\Input{A Monte Carlo search tree $\mathcal{T}$.}
	\caption{Decide($\mathcal{T}$)}
	\label{alg:decision}
	$rt \leftarrow root(\mathcal{T})$\;
	$\astate \leftarrow$ child node of $rt$ with the highest $ \rew{rt, \astate} + \vals $\;
	$\mathcal{T}\leftarrow$ the subtree of $\mathcal{T}$ rooted at $\astate$;
\end{algorithm}

\subsection{Combine Everything Together}
Finally, we combine all modules together as in Alg.~\ref{alg:MCTS} to form the MCTS framework for QCT.
Note that, to ensure the reliability of the \emph{Decision} module, 
a sufficiently large number ($\Nbp$, a predefined parameter) of \emph{Selection}, \emph{Expansion}, \emph{Simulation},  and \emph{Backpropagation}, should be performed to get a good estimation of the values of relevant states. 

Due to the stochastic nature of our algorithm, there is a negligible but still positive possibility that at certain iteration of the while loop in Alg.~\ref{alg:MCTS}, even the best child node derived from the \emph{Decision} module cannot execute any new gate. To guarantee termination in this extreme case, a 
\emph{fallback} mechanism, which has been widely used in the literature (cf. \cite{Childs}), is adopted. Specifically, if no
CNOTs have been executed after $|V|$ consecutive \emph{Decisions} and the current root node is $(\tau, PC, LC)$, then we choose a CNOT from $\mathcal{L}_0(LC)$ with minimum swap cost with respect to $\tau$, and insert the corresponding SWAP gates to $PC$ so that progress will be made by executing this chosen CNOT. For the sake of readability, the fallback module is omitted in Alg.~\ref{alg:MCTS}.

\begin{algorithm}[ht]
	\SetKwData{Left}{left}\SetKwData{This}{this}\SetKwData{Up}{up}
	\SetKwFunction{Union}{Union}\SetKwFunction{FindCompress}{FindCompress}
	\SetKwInOut{Input}{input}\SetKwInOut{Output}{output}
\hspace*{-4mm}	\Input{An architecture graph $AG$, a logical circuit $LC$, and an initial mapping $\tau_{ini}$.}
\hspace*{-4mm}	\Output{A physical circuit satisfying the connectivity constraints in $AG$.}
	\caption{Quantum circuit transformation based on Monte Carlo tree search}
	\label{alg:MCTS}
	$PC \leftarrow$ the circuit consisting of all executable gates in $LC$ under $\tau_{ini}$\;
	$LC \leftarrow LC$ with gates in $PC$ deleted\;
	$\astate\leftarrow (\tau_{ini}, PC, LC)$\;
	$\vals, \nums \leftarrow 0$\;
	$\mathcal{T} \leftarrow$ a search tree with a single (root) node $\astate$\;

	\While{$LC(\astate)\neq \emptyset$}
	{
	    \Do{$\Nbp$ times}
	    {
	        $\astate \leftarrow$ Select($\mathcal{T}$)\; 
	        Expand($\mathcal{T},\astate$)\;

			Simulate($\mathcal{T},\astate$)\;
	       	Backpropagate($\mathcal{T},\astate$)\;
	        
	    }
	    Decide($\mathcal{T}$)\; \blue{\tcp{$\mathcal{T}$ is updated in  \emph{Decide} and \emph{Expand} modules}}
	    $\astate\leftarrow root(\mathcal{T})$\;
	}
	\Return{$PC(\astate)$}
\end{algorithm}

%
%
%
%
%

\subsection{Complexity Analysis}\label{sec:complexity}

This subsection is devoted to a rough analysis of the complexity of our algorithm. Suppose $AG = \left( {V,E} \right)$ and the input logical circuit $LC = \left( {Q,C} \right)$. 
Among the five main modules presented in subsection~\ref{sec:modules}, the most expensive ones are \emph{Selection}, \emph{Expansion}, and \emph{Simulation}. We analyse their complexity separately as follows.

\textbf{Selection}. The complexity of this module depends on the depth of the search tree. In the worst case, each of the $\Nbp$ iteration in the \textbf{do} loop of Alg.~\ref{alg:MCTS} increases the depth by 1. Taking into account the fallback introduced in the last subsection, the depth of the search tree is at most $\Nbp\cdot |V|$. As each node has at most $\left| E \right|$ children, the overall complexity for this module is $O\left( {{\Nbp}\cdot |V| \cdot \left| E \right|} \right)$.

\textbf{Expansion}. There are at most $\left| E \right|$ pertinent SWAP gates available to create new nodes, and for each new one, at most $\left| C \right|$ gates need to be checked to see whether they are executable. Thus the time complexity is $O\left( {\left| E \right|\cdot \left| C \right|} \right)$. Here $|C|$ denotes the number of gates in $C$.

\textbf{Simulation}. Computing the probability distribution in Eq.~\eqref{eq:simulation} takes time $O(|E|\cdot |V|)$. To guarantee termination, the while loop will be aborted if no gates have been executed after $\left| V \right|$ consecutive iterations.  Hence, the complexity of this module is $O(|E|\cdot |V|^2\cdot \Gsim\cdot \Nsim)$.

Finally, note that in the worst case, all gates from $C$ are executed by the fallback mechanism which is invoked after every $\left| V \right|$ iterations. Hence, the \emph{Sel-Exp-Sim-BP} modules will be run for at most $|C|\cdot |V|\cdot \Nbp$ times, and the overall time complexity of our algorithm is
\begin{equation*}
O\left(|C|\!\cdot\! |V|\! \cdot\! \Nbp\! \cdot\! |E|\! \cdot\! [\Nbp\!\cdot\! |V| + |C| + |V|^2\!\cdot\!  \Gsim \!\cdot \!\Nsim] \right),
\end{equation*}
or $O(|C|\!\cdot\!|V|\!\cdot\!|E|\!\cdot\! (|C| + |V|^2))$ when the parameters are regarded as constants.

\section{Depth Optimisation}
\label{sec:MCTS-Depth}
QPUs in the NISQ era also suffer from limited coherence time, meaning that the depth of the output physical circuit is also an important criterion for optimising the circuit transformation process. In this section, we propose MCTS-Depth, which is adapted from the MCTS-Size algorithm presented in the previous section by introducing two minor changes, to further reduce the depth of the output circuit.

Recall that in MCTS-Size we have removed all single-qubit gates because they have no effect when the QCT objective is to minimise the number of inserted SWAP gates. However, as shown in Example~\ref{ex:depth}, this is not the case as far as circuit depth is concerned. In this paper, we adopt a simple strategy to deal with these gates: whenever an executable CNOT $g$ is removed from the logical circuit and added to the physical circuit in \emph{Expansion}, all single-qubit gates after $g$ and before any other CNOT that directly depends on $g$ will be greedily added to the physical circuit. 


\begin{figure}[t]
	\centering
	\includegraphics[width=0.8\textwidth]{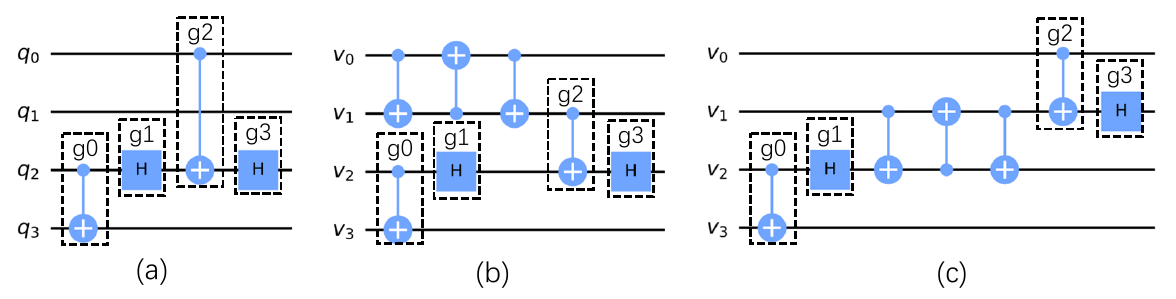}
	\caption{An input quantum circuit (a)
	and two functionally equivalent physical circuits, (b) and (c), that are executable on IBM Q20 with the naive initial mapping. }
	\label{fig:ex_depth}
\end{figure}

\begin{example}
\label{ex:depth}
    Suppose the quantum (logical) circuit to be transformed by MCTS-Depth is specified as in Fig.~\ref{fig:ex_depth}(a). Assume that  the target QPU is IBM Q20 and we take the initial mapping to be the naive one. As the CNOT $g_0$ is directly executable, $g_0$ and the single-qubit gate $g_1$ are immediately added to the physical circuit. To make $g_2$ executable, we can insert a SWAP either between physical qubits $v_0$ and $v_1$ (cf. Fig.~\ref{fig:ex_depth}(b)) or between $v_1$ and $v_2$ (cf. Fig.~\ref{fig:ex_depth}(c)). The depth overhead brought by adding $\swap{v_0}{v_1}$ and $\swap{v_1}{v_2}$ are, respectively, 1 and 3, after decomposing each SWAP into 3 CNOTs.
\end{example}

As shown in Example~\ref{ex:depth}, different SWAP gates may incur different depth overheads. Let $\astate{}=(\tau, PC, LC)$ be the current state and $\astate'=(\tau', PC', LC')$ the child state corresponding to some SWAP.  As each SWAP is implemented as three consecutive CNOTs, the depth overhead, written $\depth{\astate'}$, is an integer between 0 and 3. 
That is, a SWAP may incur 0, 1, 2, or 3 extra layers. The precise value of $\depth{\astate'}$ is calculated as the depth difference of $PC'$ and $PC$, where the executed single-qubit gates are properly added back.
\blue{Again, we note that, in practical implementation, the depth information is stored in the form of a tuple with $|V|$ elements all initiated in 0. When a gate, say $\cnot{v_i}{v_j}$, is added to the circuit, the tuple will be updated by changing both of its $i$- and $j$-th elements to $x+1$, where $x$ is the larger original value between those two elements. In addition, the depth of the corresponding circuit is exactly the maximum value in the tuple.}

Apparently, we prefer SWAPs with smaller $\depth{\astate'}$. This motivates us to replace the discount factor $\gamma$ in Eq.~\eqref{eq:BP} for MCTS-Size with $\gamma^{\depth{\astate'}}$ and obtain the following value-update rule for MCTS-Depth:
\begin{equation}
\label{eq:BP_depth}
    \vals  \leftarrow \max\{\vals,\ \discount^{\depth{\astate'}}\! \cdot \! [\rew{\astate, \astate'} + \val{\astate'}]\}
\end{equation}

Another modification is applied to the definition of the initial long-term value of a state $\astate{}$ in the simulation process, given in Eq.~\eqref{eq:simu_score}, where it uses $N$, the minimal number of SWAP gates required during all $\Nsim$ simulations, as an important index. Apparently, in order to reduce depth, it is more meaningful to replace $N$ with $M$, the minimal depth overhead of all $\Nsim$ simulations. That is, in MCTS-Depth, Eq.~\eqref{eq:simu_score} is replaced with
\begin{equation}
\label{eq:simu_score_depth}
\vals = \discount^{M/2} \cdot \Gsim
\end{equation}
and the second last line of  
Alg.~\ref{alg:simulation} is replaced with 
$$\vals \leftarrow \discount^{M/2} \cdot \Gsim,$$
\blue{where for the same reason as that in Eq.~\eqref{eq:simu_score} the exponent is taken as $M/2$ instead of $M$.}

It is clear that these modifications do not affect the complexity analysis given in Sec.~\ref{sec:complexity}.



\section{Incorporating Remote CNOT}
\label{sec:remote_cnot}

In above, we have seen how a circuit can be transformed by inserting SWAPs. This is sometimes not desirable as the mapping will change with the inserted SWAPs (cf. Example~\ref{ex:remotecnot} below). Several transformers (including the current version of \tket) suggest using remote CNOT operations (also known as bridge gates) to execute CNOT gates whose two qubits in the current mapping are not neighbours (i.e., remote). In this section, we show how remote CNOTs can be incorporated into our MCTS-based algorithms. 

\begin{figure}[t]
	\centering
	\includegraphics[width=0.35\textwidth]{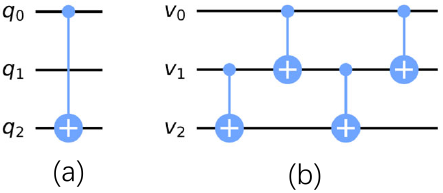}
	\caption{{Take $AG$ to be that of IBM Q20 and the initial mapping $\tau$ be naive. (a) A logical circuit with only one gate $\cnot{q_0}{q_}$. 
	(b) A remote CNOT implementation of  $\cnot{q_0}{q_2}$.}
	}
	\label{fig:ex_remote}
\end{figure}

Let $\tau$ be the current mapping and $g = \cnot{q}{q'}$. If the two physical qubits $\tau(q)$ and $\tau(q')$ are not neighbours in the target $AG$, we may replace $g$ with a sequence of CNOT gates, written $\remote{\tau}{g}$, which are executable and functionally equivalent to $g$.   
Fig.~\ref{fig:ex_remote}(b) shows the special case when the distance of $\tau(q)$ and  $\tau(q')$ in $AG$ is 2. More general construction can be found in \cite{Ud1}. 

\begin{example}
\label{ex:remotecnot}
Consider the circuit shown in Fig.~\ref{fig:fig_cir_DG} (left). Except $g_0$, every CNOT in the circuit can be executed by the naive mapping. If only SWAPs are allowed, we need to insert a SWAP to execute $g_0$. As a consequence, the mapping is changed and at least one of the other CNOTs are not executable and we need to insert another SWAP, which results in a size overhead of at least six! However, $g_0$ can be executed by implementing it as a remote CNOT depicted in Fig.~\ref{fig:ex_remote}(b) and, after that, the other CNOTs can be immediately executed, which gives an overhead of three! 
\end{example}

To extend our MCTS algorithms with remote CNOT, we need only modify \emph{Expansion} and \emph{Backpropagation}. Starting from a state/node $\astate = (\tau, PC, LC)$, besides all relevant SWAPs as used in Alg.~\ref{alg:expansion}, we also consider all $g = \cnot{q}{q'}$ if $g$ is in the first layer of $LC$ and the distance of $\tau(q)$ and $\tau(q')$ in AG is between 2 and some fixed integer $d$. We then may replace $g$ with a CNOT sequence $\remote{\tau}{g}$ along the shortest path connecting $\tau(q)$ and $\tau(q')$ if desirable. 
As remote CNOTs and SWAPs may incur different size and depth overheads, a modified value-update rule like Eq.~\eqref{eq:BP_depth} is used during  \emph{Backpropagation} if 
$\astate' = (\tau', PC', LC')$ is the child node of $\astate$ derived by a remote CNOT implementation of $g$. More precisely, for MCTS-Size, $\depth{\astate'}$ is defined as $(|\remote{\tau'}{g}|-1)/3$. The intuition behind this is that the sub-circuit $\remote{\tau'}{g}$ we used to replace $g$ brings a size overhead of $|\remote{\tau'}{g}|-1$ in terms of CNOTs, which translates to $(|\remote{\tau'}{g}|-1)/3$ in terms of SWAPs. For MCTS-Depth, $\depth{\astate'}$ is set as the depth overhead brought by adding gates in $\remote{\tau'}{g}$ to the physical circuit in state $\astate'$.

To conclude this section, we point out that the remote CNOT approach does not always give better result than the SWAP-based approach. This is because inserting SWAPs changes the mapping, which is sometimes desirable as the new mapping may execute more later CNOTs. Consider again the circuit in Fig.~\ref{fig:fig_cir_DG} (left). If the CNOT gate $g_3$ were applied on $q_0$ and $q_2$, then inserting  $\swap{v_0}{v_1}$ (i.e. three CNOTs) suffices to solve all gates in the circuit, while 
remote implementation of both $g_0$ and $g_3$ would introduce an overhead of six CNOTs.
In practice, however, it might not be easy to decide which approach is preferable. Thus we provide both of them as possible choices\blue{: The user may decide if s/he wants to use remote CNOT together with SWAPs as legal MCTS actions when calling our QCT algorithms.} In Sec.~\ref{sec:benchmark} we will evaluate its impact for two AGs.

\section{Implementation and Evaluation}
\label{sec:benchmark}
To evaluate our approach, we compare it with five state-of-the-art algorithms (cf. Sec.~\ref{sec:alg_comp}).
As the choice of initial mappings may sometimes influence the performance of QCT algorithms, to make a fair comparison, we always take the same initial mappings in their original design if available.
We use Python as our main programming language and IBM Qiskit \cite{qiskitIBM} as the auxiliary environment to implement
our algorithms. 
For efficiency, the \emph{Simulation} module is implemented in C++. All experimental results reported here are obtained by choosing the best one from five trials. Note that we only provide summarised results here. \blue{Readers are referred to the GitHub repository\footnote{https://github.com/BensonZhou1991/Circuit-Transformation-via-Monte-Carlo-Tree-Search} for detailed empirical results together with source code of our algorithms and benchmarks used in the experiments.}
 
\subsection{Benchmarks and Compared State-of-the-Art QCT Algorithms}
\label{sec:alg_comp}
In our evaluation, we selected a set of 114 benchmark circuits, with a sum of 554,497 gates (including 248,553 CNOTs) and a sum of 303,469 depths, \blue{which, taken from \cite{Astar}, were  published by IBM as part of the 2018 QISKit Developer Challenge\footnote{https://www.ibm.com/blogs/research/2018/08/winners-qiskit-developer-challenge/} and have been} widely used in evaluating circuit transformation algorithms by, e.g., \cite{CowtanRouting,SubgraphIsomorphism,SAHS,LiUCSB}. \blue{In the following, we write $\B{114}$ for this benchmark set.}

Although circuits in $\B{114}$ are widely used, not all of them are directly relevant to quantum computing. To evaluate the proposed algorithms on `real' quantum circuits, we also extracted a set of 173 quantum circuits, written $\B{real}$, from the quantum algorithm library in Qiskit. Circuits in $\B{real}$ have a sum of 603,654 gates  (including  260,589 CNOTs) and a sum of 413,734 depths.


\blue{The QCT algorithms to be compared with our MCTS algorithms include  \tket\ (version 0.17.0\footnote{https://cqcl.github.io/pytket/build/html/index.html}) \cite{CowtanRouting}, SAHS \cite{SAHS}, FiDLS \cite{SubgraphIsomorphism}, Qiskit (version 0.33.0) \cite{qiskitIBM} and SABRE \cite{LiUCSB}, which are state-of-the-art algorithms for quantum circuit transformation. While we didn't include Cirq\footnote{https://quantumai.google/cirq} in our comparison, it was found  in  \cite{tan_benchmark} that Cirq is less efficient than \tket\ and Qiskit.} 
For fair and pure comparison of the routing abilities, we also disabled the postmapping
optimisation of \tket. 



\subsection{Parameter Determination}
\begin{figure*}[t]
	\centering
	\subfigure[]{
		\begin{minipage}[t]{0.31\linewidth}
			\centering
			\includegraphics[width=1\linewidth]{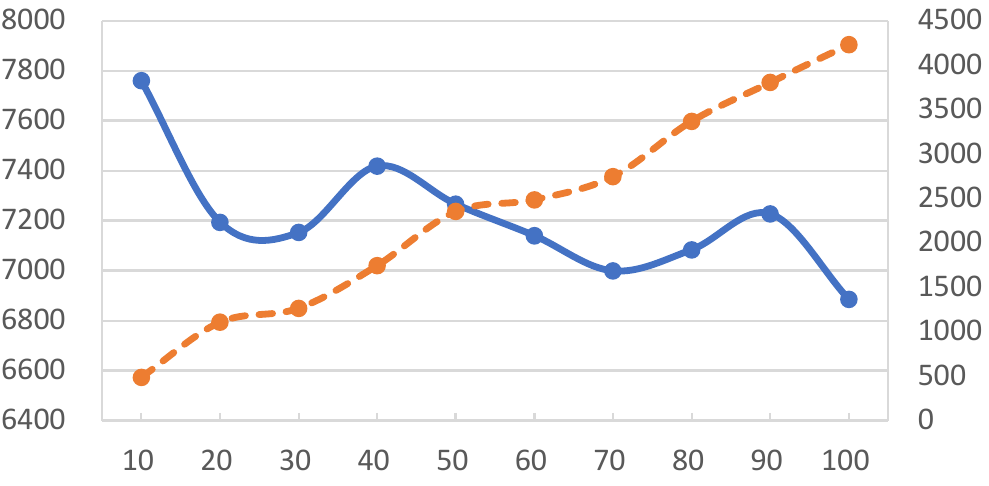}
		\end{minipage}
	}
	\subfigure[]{
		\begin{minipage}[t]{0.31\linewidth}
			\centering
			\includegraphics[width=1\linewidth]{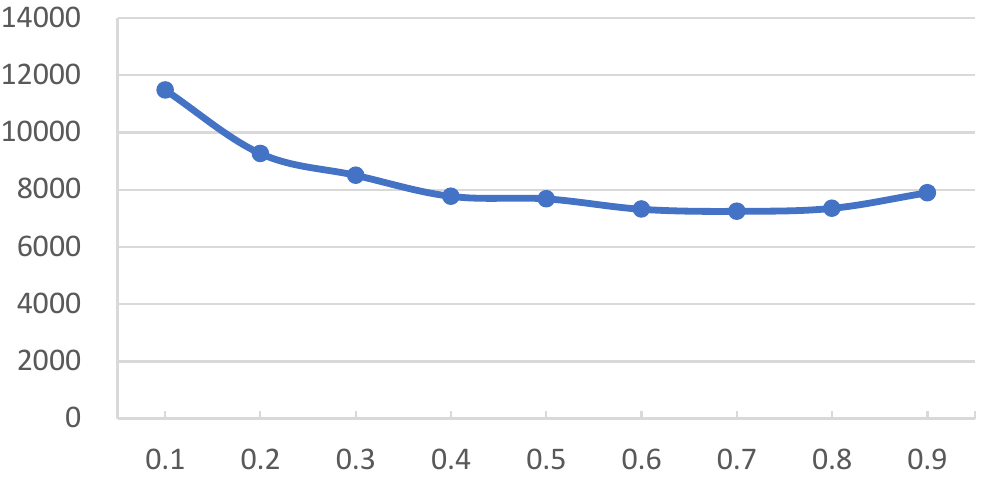}
		\end{minipage}
	}
	\subfigure[]{
		\begin{minipage}[t]{0.31\linewidth}
			\centering
			\includegraphics[width=1\linewidth]{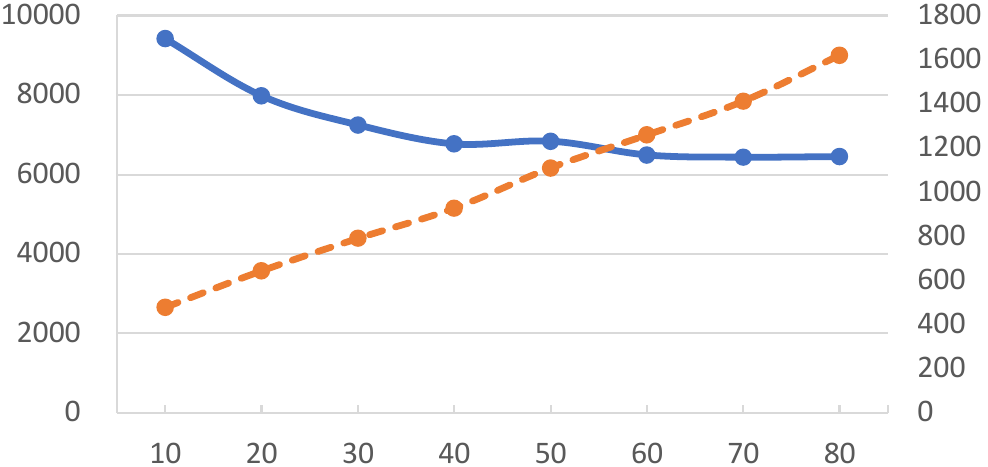}
		\end{minipage}
	}
	\centering
	\caption{Evaluation of the performance of MCTS-Size on IBM Q20 for different parameter settings: (a) $\Nbp$, (b) $\discount$, and (c) $\Gsim$, where the vertical axes on the left and right in each sub-figure represent the aggregated number of added CNOTs in output circuits (blue lines) and running time (seconds, orange dashed lines), respectively. All data is aggregated from the small benchmark set $\B{11}$	and initial mappings are the same as those used in SAHS \cite{SAHS}.}
	\label{fig:parameter}
\end{figure*}

Our MCTS-based QCT algorithms have a couple of parameters to {be} determined before actual running:
\begin{itemize}
\item $\Nbp$ (repeated times for the \emph{Sel-Exp-Sim-BP} modules before each \emph{Decision}), 
\item $\cexpl$ (the exploration parameter used in Eq.~\eqref{eq:selection}), 
\item $\Gsim$ (the size of sub-circuit used in simulation), 
\item $\Nsim$ (the number of simulations),
\item $\discount$ (the discount ratio), and
\item $d$ (the maximum distance allowed for remote CNOT).
\end{itemize}
\blue{To help determine these parameters for IBM Q20, we selected a small subset of 11 typical circuits from $\B{114}$ each of which involves 10-15 qubits and has 1000-6000 CNOTs. This benchmark set, written as $\B{11}$, contains in total 26,676 CNOTs.}
\blue{Fig.~\ref{fig:parameter} depicts the dependency of the size of the final physical circuits (left vertical axis) and the running time (right vertical axis) on different parameter settings.}
\blue{One may note that the performance in Fig.~\ref{fig:parameter}(a) does not improve steadily with the increasing of $\Nbp$. This is perhaps due to that those data are derived by running our algorithm for five times and keeping only the best results.}

To get a good balance between performance and running time, we empirically set\\
\centerline{$\Nbp = \cexpl = 20$, $\Gsim = 30$, $\Nsim = 500$, and $\discount = 0.7$.}
We adopt the same parameter setting for the other proposed algorithms, and {set $d=2$ for MCTS-Size and MCTS-Depth.}

\begin{figure}[t]
	\centering
	\subfigure[]{
		\begin{minipage}[t]{0.45\linewidth}
			\centering
			\includegraphics[width=1\linewidth]{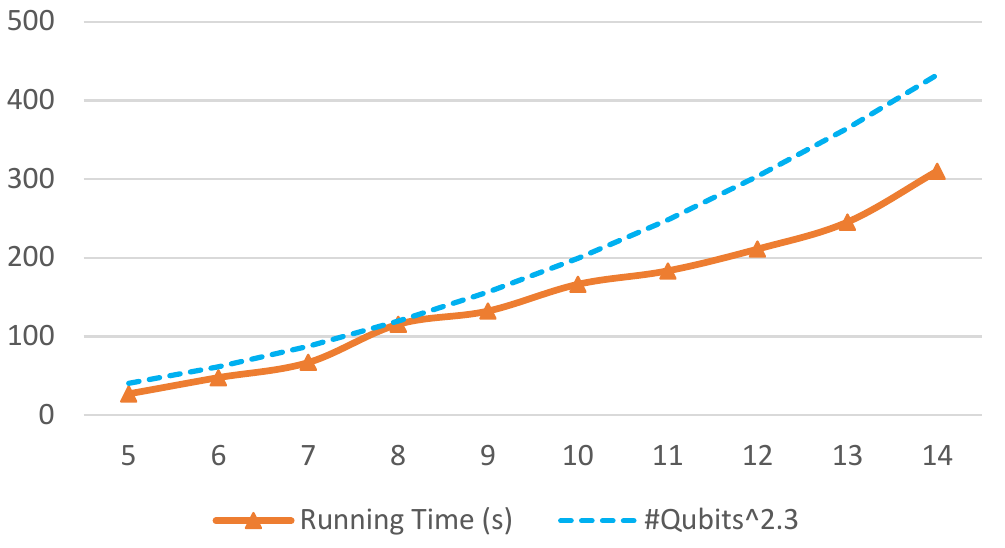}
		\end{minipage}
	}
	\subfigure[]{
		\begin{minipage}[t]{0.45\linewidth}
			\centering
			\includegraphics[width=1\linewidth]{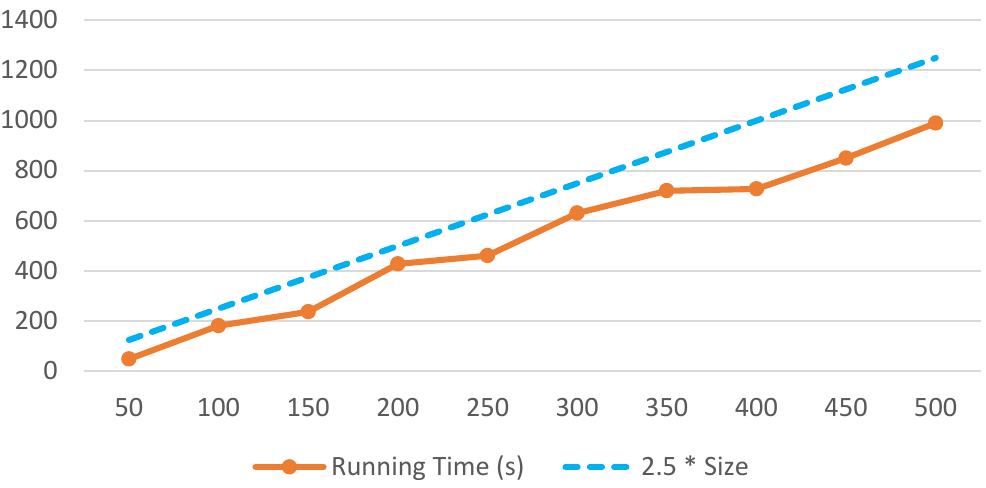}
		\end{minipage}
	}
	
	\centering
	\caption{Evaluations of the performance of MCTS-Size on a hypothetical AG Grid $4\times 5$ by comparing the average running time (seconds, orange line) with (a): number of logical qubits, and (b): number of CNOTs of the input circuits, where each circuit in (a) contains 500 CNOTs and each circuit in (b) has 20 qubits, and each data point denotes the average value of 10 randomly generated quantum circuits with naive initial mappings.
	} 
	\label{fig:complexity}
\end{figure}

\subsection{Running Time and Search Depth}

We have shown in Sec.~\ref{sec:complexity} that our algorithm runs in time polynomial in all relevant parameters.
To further demonstrate the running time in practice, we randomly generate two sets of 10 quantum circuits. In one set, each circuit has 500 CNOTs, and the number of logical qubits ranges from 5 to 14. In the other, each circuit has 20 qubits, and the number of CNOTs ranges from 50 to 500. 
We transform all these circuits via MCTS-Size on a hypothetical AG Grid $4\times 5$, and record the average running time for each circuit set. As shown in Fig.~\ref{fig:complexity}, the real time cost is roughly the 2.3th power in the number of qubits and linear (with slope being about 2.5) in the number of CNOTs, indicating that our algorithm is practically scalable.

\begin{figure}[t]
	\centering
	\includegraphics[width=\figw\textwidth]{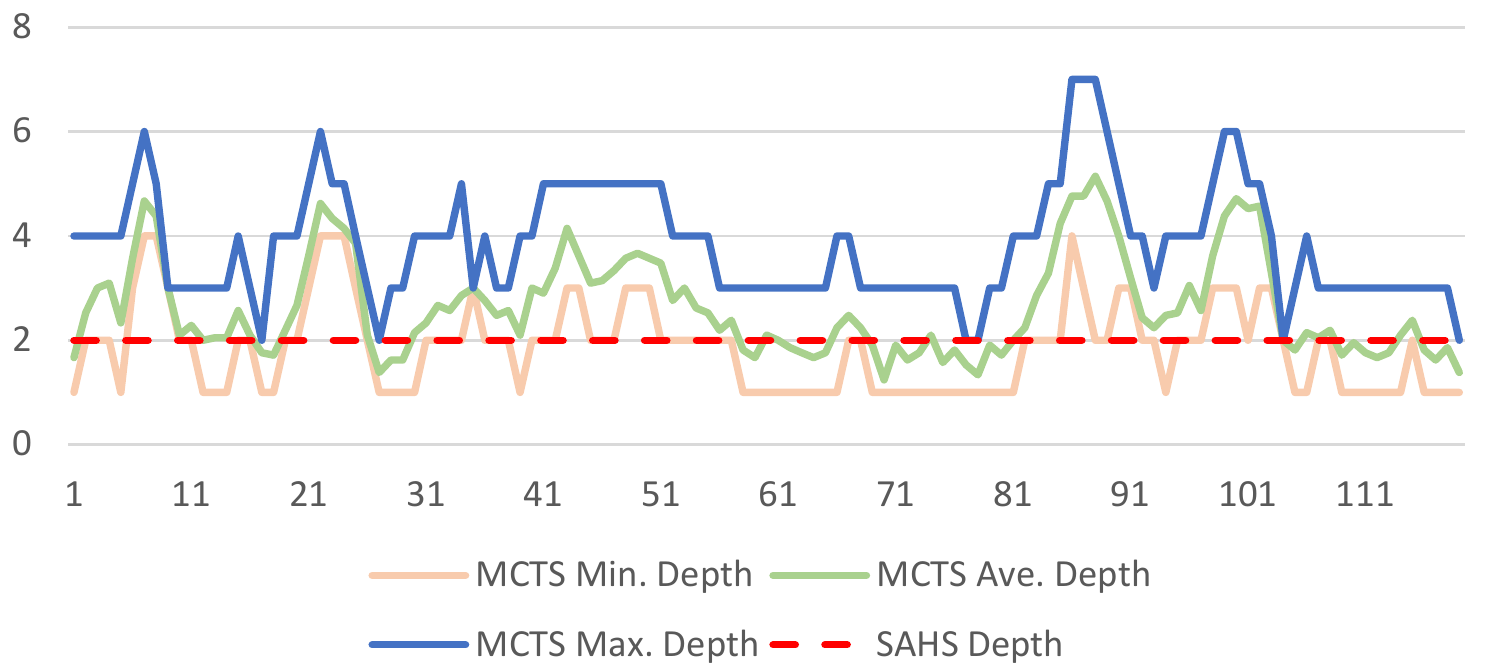}
	\caption{Search depth (for \emph{Selection} before each \emph{Decision} ) of {MCTS-Size} for circuit `misex1\_241'. The horizontal and vertical axis represent rounds for \emph{Decision} and search depth, respectively. \blue{Note that here the search depth of SAHS is fixed as 2.}
	}
	\label{fig:depth}
\end{figure}

For MCTS-Size, we also record the search depth for each {use of the} \emph{Selection} module and calculate the minimum, average, and maximum depth before each \emph{Decision} process. \blue{In SAHS \cite{SAHS}, the search depth is an adjustable parameter determining how deeply the heuristic evaluation process will look into.} As shown in Fig.~\ref{fig:depth}, the maximum search depth of MCTS-Size can easily exceed that of SAHS, which is set to 2 \blue{by default in its original implementation}. Actually, in most of the time it is more than 3, meaning that our algorithm has better ability of exploring the unknown state. 
\begin{figure}[t]
	\centering
	\includegraphics[width=\figw\textwidth]{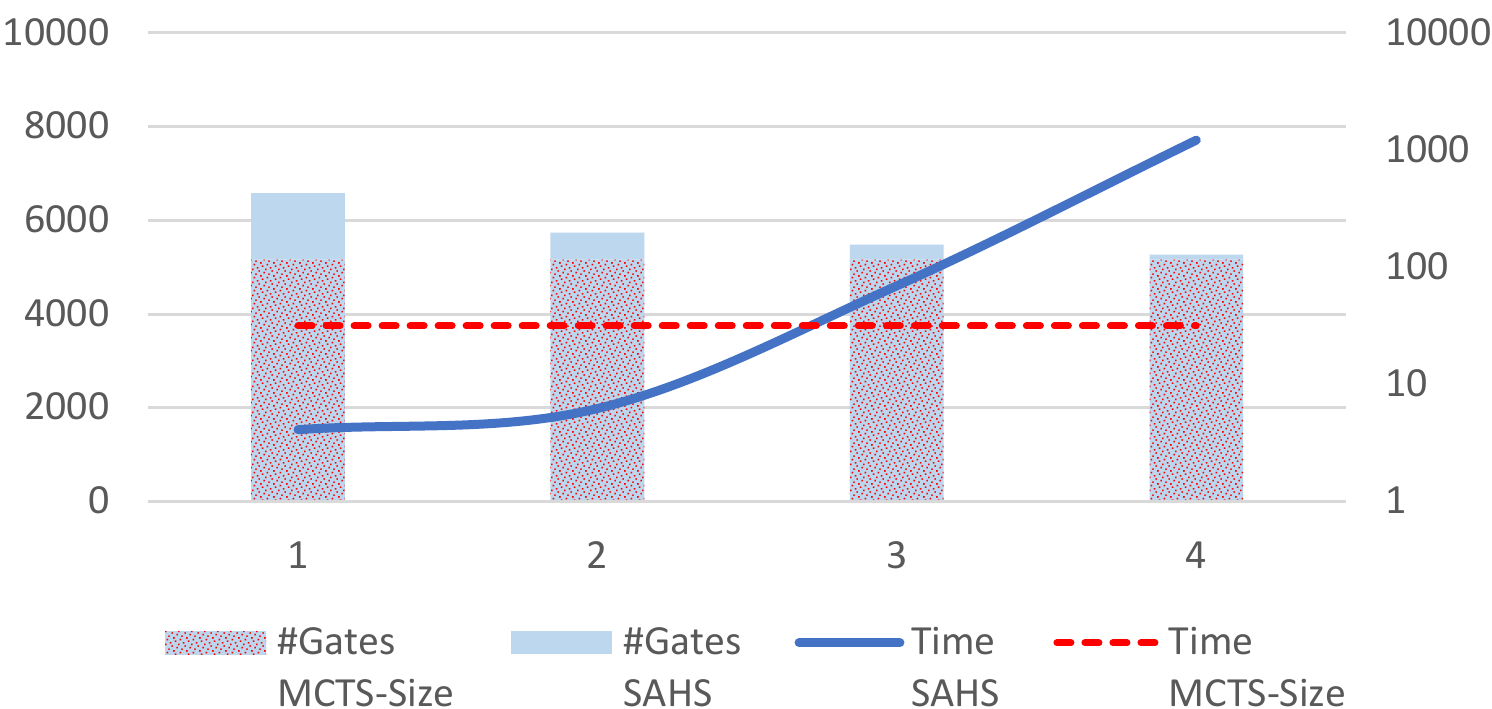}
	\caption{The benefit brought by increasing the search depth in SAHS~\cite{SAHS} for circuit `misex1\_241'. The horizontal axis and vertical axes on the left and right represent search depth, number of gates in the output circuit and running time (seconds), respectively. \blue{Note that here the search depth is only applied to SAHS and it is not adjustable in MCTS.}
	}
	\label{fig:depth_SAHS}
\end{figure}

Note that it is claimed in~\cite{SAHS} that the size of output physical circuits can be further decreased by increasing the search depth, with the cost of more time consumption. 
Fig.~\ref{fig:depth_SAHS} depicts a comparison of the output circuit size as well as the running time of MCTS-Size and SAHS on the example circuit `misex1\_241' with 4,813 gates, where the search depth of SAHS varies from 1 to 4. 
It shows that MCTS-Size outperforms SAHS in the output circuit size even when the search depth of SAHS is set to 4. However, in this case,  the running time of SAHS is over 20 minutes, while MCTS-Size only needs 31 seconds.

\subsection{Evaluation of the Performance of MCTS-Size}

Now we compare MCTS-Size with SAHS \cite{SAHS}, FiDLS \cite{SubgraphIsomorphism}, and \tket\cite{CowtanRouting} on {IBM Q20} over the 114 benchmark circuits in $\B{114}$. The results are summarised in Table~\ref{tab:summary}, where 
{columns 2 \& 3 represent aggregated numbers of added CNOTs (each SWAP is decomposed into 3 CNOTs)} obtained from other methods and MCTS-Size when using their initial mappings, respectively.
Besides, the `improvement' is defined as $(n_{comp} - n_{ours})/n_{comp}$, with $n_{comp}$ and $n_{ours}$ being the total numbers of CNOT gates added by the compared algorithm and ours, respectively, in transforming the 114 circuits. A similar definition for `Improvement' is used in the rest of the paper. 

\begin{table}[h]
	\centering
	\caption{
	\blue{Comparison of MCTS-Size with state-of-the-art algorithms using their initial mappings on the large benchmark set $\B{114}$,
	where Columns 2 \& 3 represent aggregated numbers of added CNOTs obtained from the algorithm under comparison and MCTS-Size when using their initial mappings, respectively.}
	}
	\label{tab:summary}
	\begin{tabular}{c|ccc}
		Compared Algorithm & gate overhead & \begin{tabular}[c]{@{}c@{}} gate overhead\\      MCTS-Size\end{tabular} & Improvement \\
		\hline
		SAHS & 116487 & 73758 & 36.68\% \\
		Topg. FiDLS & 107406 & 74763 & 30.39\% \\
		Wgt. FiDLS & 105645 & 75126 & 28.89\% \\
		{\tket} & 238170 & \emph{77544} &  59.24\% \\
	\end{tabular}
\end{table}

From Table~\ref{tab:summary}, we can see that MCTS-Size achieves a conspicuous improvement of 36.68\% on average when compared with SAHS (by using the same initial mappings as SAHS). In \cite{SubgraphIsomorphism},  two techniques for initial mappings, topgraph (topg.) and weighted graph (wgt.), are proposed with FiDLS. Our algorithm has a consistent improvement, 30.39\% for the topgraph initial mappings and 28.89\% for the weighted graph ones. 
As the initial mappings of \tket\ are not directly available, we use naive mappings as the initial mappings in the experiments. For fair and pure comparison of the routing abilities, we also disabled the postmapping optimisation of \tket. As can be seen from the last row of Table~\ref{tab:summary}, the gate overhead of \tket\ is above 3 times of ours (also with naive initial mappings).
It is worth noting that, on this benchmark set, MCTS-Size still performs well even with naive initial mappings; the overhead compared with its best result is only 5.13\% (77,544 vs. 73,758). 

\blue{For each compared algorithm, MCTS-Size performs at least as good as the compared algorithm on most circuits. Particularly, the performance of MCTS is consistently better than or the same as SAHS in all circuits. Furthermore, an improvement less than 5\% occurs only when the circuit is small and can be transformed by the compared algorithm with no more than 30 ancillary SWAP (or, equivalently, 90 CNOT) gates.  Hence, it can be concluded that the performance of the MCTS algorithm is stable in various input circuits and initial mappings.}

\subsection{Evaluation of the Performance of MCTS-Depth}

Now we compare MCTS-Depth with \blue{MCTS-Size, {\tket}, Qiskit, and SABRE} on \blue{the two benchmark sets} in terms of the total depth overhead. As the initial mappings used by \tket\ are not directly available, we adopt the naive mapping as the initial mapping for all these algorithms. It is surprise that on IBM Q20 the initial mappings selected by \tket\ are, on average, not better than the naive mappings (238,170 vs. 237,471 added CNOTs and 238,051 vs. 236,312 added depths). 

\blue{To evaluate the impact of the underlying topology of the architecture structure, we also run the experiments on a hypothetical grid-like QPU, called Grid $4\times 5$, which has fewer edges (31 vs. 43) than IBM Q20. We also empirically evaluated the impact of remote CNOT gates in our MCTS algorithms and \tket.} As can be seen from Table~\ref{tab:depth_benchmark}, 
MCTS-Depth is able to improve the depth performance steadily for both tested AGs when compared to MCTS-Size, which confirms the utility of our modifications in Sec.~\ref{sec:MCTS-Depth}. We note that the improvements of Alg.~X against Alg.~Y in Table~\ref{tab:depth_benchmark} are calculated as $(n_{Y}-n_X)/n_Y$ with $n_Y$ ($n_X$, resp.) being either the sum of the added CNOTs or the sum of the added depths by Alg.~Y (Alg.~X, resp.).

\subsubsection{Impact of the Topology of the Architecture Structure}

When compared with \tket, for IBM Q20, both MCTS-Size and MCTS-Depth have a great advantage, with about 75\% and 84\% improvement; for Grid $4\times 5$, however, their advantages are not so remarkable. The main reason is perhaps due to the fact that IBM Q20 supports more qubit connections (and has a smaller diameter) than Grid $4\times 5$, which enables the MCTS-based algorithms to find a good solution without going much deeper for IBM Q20. Another reason may be that \tket\ performs really well on grid-like AGs.

\subsubsection{Impact of Remote CNOT}

Next, we discuss the improvement brought by introducing remote CNOT. We empirically evaluated the impact of remote CNOT in our MCTS-based algorithms and \tket. The results are summarised in Table~\ref{tab:depth_benchmark}, where for an algorithm A, A+r denotes the algorithm with remote CNOT enabled.
For IBM Q20, the improvements brought by introducing remote CNOT are almost negligible, and sometimes even degraded. For Grid $4\times 5$, however, the improvements are quite significant. Compared to \tket, our algorithms have gained an improvement of 28\% in size (MCTS-Size+r) and an improvement of 42\% in depth (MCTS-Depth+r), while \tket+r has a 4\% improvement in size and a 14\% improvement in depth.

\subsubsection{Evaluation on Real Benchmark Circuits}
We also evaluated these algorithms on IBM Q20 and the real benchmark circuits in $\B{real}$. The results are summarised in Table~\ref{tab:quantum_benchmark}, where we can see that (i) the impact of remote CNOT is significantly negative; (ii) the improvements of both MCTS algorithms against \tket\ are still above 50\%.

\vspace*{2mm}

\blue{In addition, we also compared MCTS-Depth with the new algorithm proposed in \cite{zhang2021time}, which aims at minimising the depth of the output circuit. Evaluation on IBM Q20 and 26 benchmark circuits used in \cite{zhang2021time} shows that MCTS-Depth performs consistently better and can reduce on average the depth overhead by 67\% when compared with the algorithm in \cite{zhang2021time}.}


\begin{table}[h]
\centering
\caption{
\blue{Comparison of the proposed MCTS algorithms with \tket, Qiskit and SABRE on the benchmark set $\B{114}$ with naive initial mappings, where A+r denotes the algorithm A with remote CNOT enabled. Columns 5 \& 6 represent, respectively, the size and depth improvement compared to \tket.
}
}
\label{tab:depth_benchmark}
\begin{tabular}{cc|cccc}
AG & Method & \begin{tabular}[c]{@{}c@{}}CNOT\\      Added\end{tabular} & \begin{tabular}[c]{@{}c@{}}Depth\\      Added\end{tabular} & \begin{tabular}[c]{@{}c@{}}Size\\      Imp.\end{tabular} & \begin{tabular}[c]{@{}c@{}}Depth\\      Imp.\end{tabular} \\
\hline
\multirow{6}{*}{IBM Q20} & \tket & 237471 & 236312 & - & - \\
 & \tket+r & 273420 & 238661 & -15.14\% & -0.99\% \\
  & Qiskit & 575400 & 335133 & -142.30\% & -41.82\% \\
 & SABRE & 406365 & 382442 & -71.12\% & -61.84\% \\
 & MCTS-Size & 79743 & 58602 & 66.42\% & 75.20\% \\
 & MCTS-Size+r & 79275 & 59857 & 66.62\% & 74.67\% \\
 & MCTS-Depth & 151530 & 37794 & 36.19\% & 84.01\% \\
 & MCTS-Depth+r & 147972 & 37138 & 37.69\% & 84.28\% \\
 \hline
\multirow{6}{*}{Grid $4\times 5$ } & \tket & 383409 & 368109 & - & - \\
 & \tket+r & 367815 & 314148 & 4.07\% & 14.66\% \\
 & Qiskit & 641616 & 456261 & -67.35\% & -23.95\% \\
 & SABRE & 596679 & 517354 & -55.62\% & -40.54\% \\
 & MCTS-Size & 356091 & 359240 & 7.13\% & 2.41\% \\
 & MCTS-Size+r & 275274 & 263811 & 28.20\% & 28.33\% \\
 & MCTS-Depth & 514311 & 292050 & -34.14\% & 20.66\% \\
 & MCTS-Depth+r & 392349 & 211421 & -2.33\% & 42.57\%\\
\end{tabular}
\end{table}



\begin{table}[]
\centering
\caption{
\blue{Comparison of the proposed MCTS algorithms with \tket, Qiskit, and SABRE over quantum circuits in the benchmark set $\B{real}$ on IBM Q20.}
}
\label{tab:quantum_benchmark}
\begin{tabular}{cc|cccc}
AG & Method & \begin{tabular}[c]{@{}c@{}}CNOT\\      Added\end{tabular} & \begin{tabular}[c]{@{}c@{}}Depth\\      Added\end{tabular} & \begin{tabular}[c]{@{}c@{}}Size\\      Imp.\end{tabular} & \begin{tabular}[c]{@{}c@{}}Depth\\      Imp.\end{tabular} \\
\hline
\multirow{6}{*}{IBM Q20} & \tket & 101304 & 91730 & - & - \\
 & \tket+r & 156774 & 121734 & -54.76\% & -32.71\% \\
 & Qiskit & 799818 & 514119 & -93.64\% & -9.44\% \\
 & SABRE & 802737 & 597933 & -96.52\% & -100.81\% \\
 & MCTS-Size & 41010 & 41826 & 59.52\% & 54.40\% \\
 & MCTS-Size+r & 38607 & 49012 & 61.89\% & 46.57\% \\
 & MCTS-Depth & 64992 & 27237 & 35.84\% & 70.31\% \\
 & MCTS-Depth+r & 60333 & 31300 & 40.44\% & 65.88\% 
\end{tabular}
\end{table}

\begin{table}[h]
\centering
\caption{
\blue{Comparison of the proposed MCTS algorithms with \tket, Qiskit and SABRE on two benchmark sets $\B{114}$ and $\B{ran}$ with naive initial mappings on IBM Rochester.}
}
\label{tab:large_ag_benchmark}
\begin{tabular}{cc|cccc}
Benchmark & Method & \begin{tabular}[c]{@{}c@{}}CNOT\\      Added\end{tabular} & \begin{tabular}[c]{@{}c@{}}Depth\\      Added\end{tabular} & \begin{tabular}[c]{@{}c@{}}Size\\      Imp.\end{tabular} & \begin{tabular}[c]{@{}c@{}}Depth\\      Imp.\end{tabular} \\
\hline
\multirow{8}{*}{$\B{114}$} & \tket & 575418 & 560479 & - & - \\
 & \tket+r & 495687 & 429980 & 13.86\% & 23.28\% \\
 & Qiskit & 882606 & 590607 & -53.39\% & -5.38\% \\
 & SABRE & 844923 & 681685 & -46.84\% & -21.63\% \\
 & MCTS-Size & 563643 & 560205 & 2.05\% & 0.05\% \\
 & MCTS-Size+r & 970678	& 375384 & 16.04\% & 33.02\% \\
 & MCTS-Depth & 416181 & 467698 & -37.13\% & 16.55\% \\
 & MCTS-Depth+r & 513771 & 310220 & 10.71\% & 44.65\% \\
 \hline
\multirow{8}{*}{$\B{ran}$ } & \tket & 29526 & 7321 & - & - \\
 & \tket+r & 29301 & 6905 & 0.76\% & 5.68\% \\
 & Qiskit & 36296 & 5551 & -16.16\% & 24.18\% \\
 & SABRE & 30921 & 5456 & -4.7\% & 25.47\% \\
 & MCTS-Size & 26427 & 7050 & 10.50\% & 3.70\% \\
 & MCTS-Size+r & 26097 & 7167 & 11.61\% & 2.10\% \\
 & MCTS-Depth & 30378 & 3445 & -2.89\% & 52.94\% \\
 & MCTS-Depth+r & 29481 & 3491 & 0.15\% & 52.32\%
\end{tabular}
\end{table}

\begin{table}[h]
\centering
\caption{
\blue{Comparison of the proposed MCTS algorithms with \tket, Qiskit, and SABRE on two benchmark sets $\B{114}$ and $\B{ran}$ with naive initial mappings on Google Sycamore.}
}
\label{tab:large_ag_google_benchmark}
\begin{tabular}{cc|cccc}
Benchmark & Method & \begin{tabular}[c]{@{}c@{}}CNOT\\      Added\end{tabular} & \begin{tabular}[c]{@{}c@{}}Depth\\      Added\end{tabular} & \begin{tabular}[c]{@{}c@{}}Size\\      Imp.\end{tabular} & \begin{tabular}[c]{@{}c@{}}Depth\\      Imp.\end{tabular} \\
\hline
\multirow{8}{*}{$\B{114}$} & \tket & 391923 & 378233 & - & - \\
 & \tket+r & 370068 & 358354 & 5.58\% & 5.26\% \\
 & Qiskit & 625014 & 446666 & -59.47\% & -18.09\% \\
 & SABRE & 582933 & 509737 & -48.74\% & -34.77\% \\
 & MCTS-Size & 365823 & 670279 & 6.66\% & 3.02\% \\
 & MCTS-Size+r & 291732	& 366810 & 21.17\% & 26.24\% \\
 & MCTS-Depth & 471309 & 304140 & -20.26\% & 19.59\% \\
 & MCTS-Depth+r & 365940 & 213782 & 6.63\% & 43.48\% \\
 \hline
\multirow{8}{*}{$\B{ran}$ } & \tket & 15837 & 4146 & - & - \\
 & \tket+r & 15813 & 4072 & 0.15\% & 1.78\% \\
 & Qiskit &  17967&  4062& -13.45\% & 2.03\% \\
 & SABRE & 18030 & 3877 & -13.85\% & 6.49\% \\
 & MCTS-Size & 14328 & 4585 & 9.53\% & -10.59\% \\
 & MCTS-Size+r & 14148 & 4558 & 10.66\% & -9.94\% \\
 & MCTS-Depth & 16155 & 2280 & -2.01\% & 45.01\% \\
 & MCTS-Depth+r & 15765 & 2254 & 0.45\% & 45.63\%
\end{tabular}
\end{table}

\subsection{\blue{Evaluation on Large Architectural Graphs}}
\label{sec:exp_more_ag}


IBM Rochester and Google Sycamore, both having 53 qubits (cf. Fig.~\ref{fig:AG}), are two state-of-the-art QPUs. It is natural to ask if our MCTS-based algorithms still have superior performance on these QPUs. Empirical evaluations show that the MCTS algorithms as presented above do not perform significantly better than {\tket}. However, if we replace the original simulation module as described in Alg.~\ref{alg:simulation} with a simple deterministic heuristic strategy, then the MCTS algorithms also demonstrate superior performance on Sycamore and Rochester.

\subsubsection{A Deterministic Strategy for Simulation}
Recall that the goal of Simulation is to obtain a simulated score as the initial long-term value for a state. We introduce a deterministic heuristic strategy to replace the original Simulation module described in Alg.~\ref{alg:simulation}.
%
%
Specifically, when simulation is requested in a state $\astate$ with a parent state, say $\astate'$, we extract the first $\Lsim$ (a predefined parameter, fixed as 4 in our implementation) layers of gates in the logical circuit in the parent state $\astate'$, and then calculate
the initial long-term value of $\astate$  as
\begin{equation}
\label{eq:deterministic_heuristic}
\vals = 
\sum\limits_{i = 0}^{\Lsim} {\left[\left( {\sum\limits_{g \in \mathcal{L}_i(LC(\astate'))} \scost{g}{\tau'} -  
\sum\limits_{g \in \mathcal{L}_i(LC(\astate'))} \scost{g}{\tau}} \right)\cdot \discount^{i+1}\right]} 
\end{equation}
where the notations are identical to that in Eq.~\eqref{eq:sim_if}. This heuristic first calculates, for each layer in the extracted logical circuit, its score defined as the total distance reduction among all CNOTs brought by the inserted SWAP corresponding to $\astate$ and then aggregates these scores with a discount factor $\gamma$ (predefined and set as 0.7 in the experiments).

\subsubsection{Evaluation of the MCTS Algorithms with the New Simulation Module}
To demonstrate the efficacy of our MCTS-based algorithms on QPUs with large qubit numbers, experiments are also done on IBM Rochester and Google Sycamore. 
We then make comparisons with \tket, Qiskit, and SABRE. 
The results are summarised in Table~\ref{tab:large_ag_benchmark}, for IBM Rochester, and Table~\ref{tab:large_ag_google_benchmark}, for Sycamore, from which we can see
that our MCTS-based algorithms still consistently outperform those industrial-level algorithms, especially when the target is circuit depth.

\begin{figure}[t]
	\centering
	\includegraphics[width=\figw\textwidth]{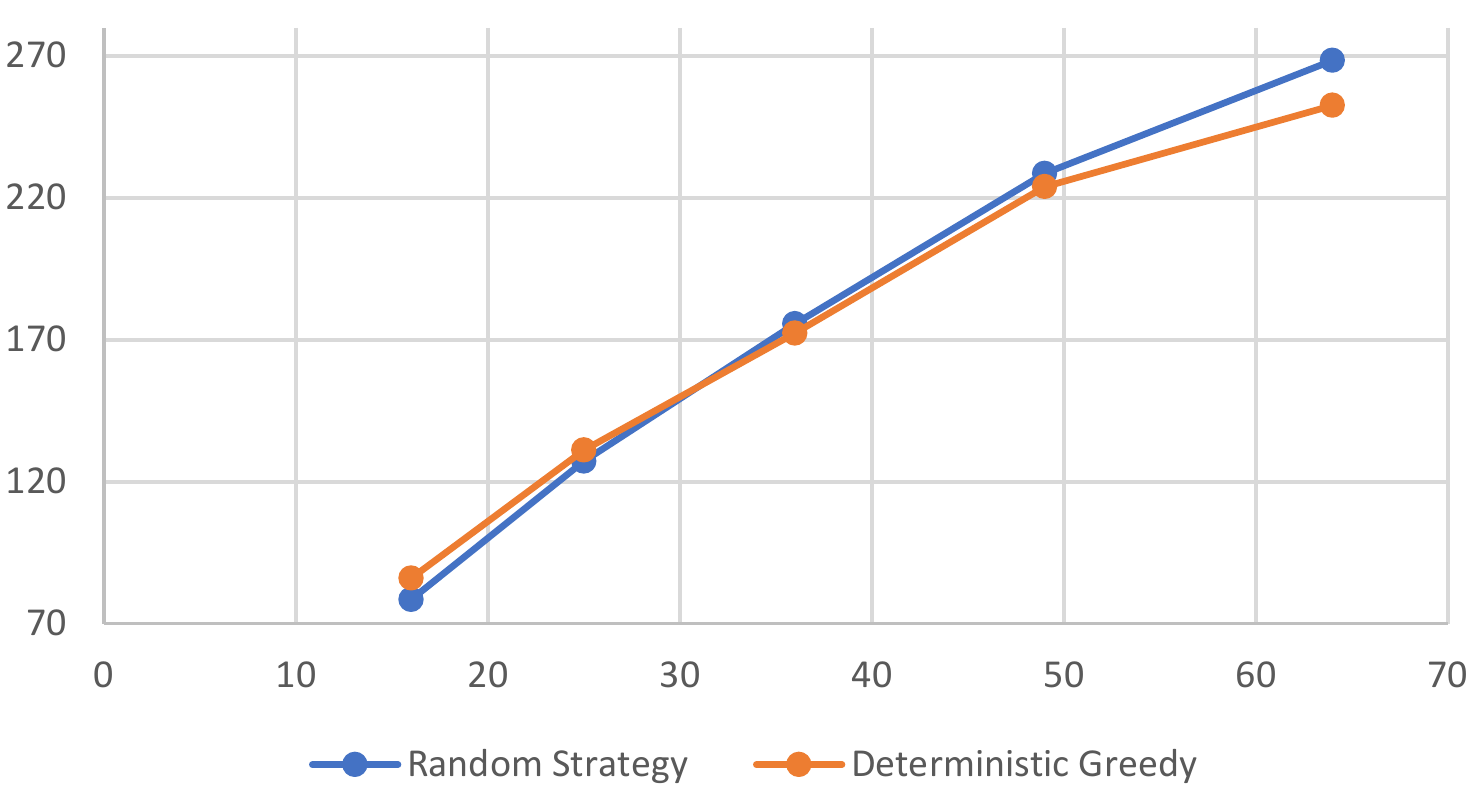}
	\caption{The average numbers of added ancillary CNOT gates derived by two native QCT algorithms (vertical axis) vs. numbers of qubits in tested AGs.}
	\label{fig:sim_compare}
\end{figure}

To explain why the deterministic strategy performs better than the random one on large AGs, a series of experiments are devised by introducing two \emph{kernel} QCT algorithms are designed: The first one utilises the random simulation strategy directly to transform the input circuit while the second adopts the deterministic heuristic in Eq.~\eqref{eq:deterministic_heuristic}. 
Then we compare the performance of the two kernel algorithms on five hypothetical AGs: Grid $4\times 4$, $5\times 5$, $6\times 6$, $7\times 7$ and $8\times 8$. For each AG, we generate 20 logical circuits each consisting of 30 randomly placed CNOT gates and then transform them to physical ones by the two native algorithms, respectively. For each AG, we record the average number of added ancillary CNOT gates among its 20 transformed physical circuits. As depicted in Fig.~\ref{fig:sim_compare}, the algorithm with the deterministic greedy strategy performs better than that with the random strategy on AGs with more than 36 qubits, indicating the rationality for adopting the new Simulation strategy for large AGs.

\section{Conclusion}
\label{sec:conclusion}
In this paper, an MCTS framework is proposed for the quantum circuit transformation problem, which aims at minimising either the size or the depth overhead to transform an ideal logical circuit to a physical one executable on a QPU with connectivity constraints. 
For this purpose, a scoring mechanism (cf. Eq.s~\ref{eq:BP} and \ref{eq:BP_depth}) is designed which takes into account both the short-term reward of introducing a SWAP 
and a long-term value obtained by random simulations. Furthermore, when backpropagating rewards collected by states to their ancestors, a discount factor is introduced to guide the algorithm towards a 
{cheapest} path to a goal state. The MCTS-based algorithms, viz., MCTS-Size and MCTS-Depth, run in polynomial time with respect to all relevant parameters. With {six} parameters,  they are very flexible in meeting different optimisation objectives, can stop whenever a preassigned resource limit is reached, and search much deeper than existing algorithms. Empirical results on extensive
realistic circuits on IBM Q20, Rochester, and Google Sycamore confirmed that MCTS-Size (MCTS-Depth, resp.) can reduce, on average,
the CNOT (depth, resp.) overhead  by as high as 75\% (84\%, resp.) when compared with \tket, an industrial level product. 



When designing QCT algorithms, we assume that logical circuits are transformed into executable physical circuits by inserting ancillary CNOTs stepwise. 
\blue{Besides CNOTs, several other tools may be considered in our MCTS-based algorithms. For example, the algorithm in \cite{gheorghiu2020reducing} tries to `re-compile' part of or the whole logical circuit to make the CNOTs in the newly compiled one satisfying the connectivity constraints; gate commutation rules are used in \cite{DG} to further simplify the transformed circuit; and quantum teleportation is introduced in \cite{hillmich2021teleportation} as a complementary method for the transforming process.}
Combination of these approaches may generate functional equivalent physical circuits with lower depth.


Recently, Tan and Cui \cite{tan_benchmark} proposed a random circuit library QUEKO for evaluating the optimality of quantum circuit transformers, which contains circuits with known optimal depth overhead. We evaluated MCTS-Depth on QUEKO and the results show a total score of {2.88} (meaning the ratio of the total depths of the output and input circuits), which is, though better than that of \tket\ (3.76), is still too far from the optimal ratio, viz. 1. This is partially due to that we used naive initial mappings, instead of the optimal mappings, e.g., those found by subgraph isomorphism \cite{SubgraphIsomorphism}. On the other hand, it suggests that there is still much room to improve the implementation of our algorithms. This is the first problem we intend to attack for future studies. Second, parameters presented in our algorithms are QPU-dependent, and a careful study of their correlation may provide a better insight on how to choose them in practice. Third, more objectives, e.g., fidelity and error rate, should be included in evaluating the quality of output physical circuits. Last but not least, it is promising to develop a parallelised implementation of our MCTS-based algorithms in a multi-thread way, where hundreds or thousands computational processes can run in parallel and share the same memory. The success implementation of this parallelised MCTS framework could help us get even better results (by going deeper) more quickly.

\section{Acknowledgements}
We thank the reviewers for their very helpful comments and suggestions. This work was supported by the National Key R\&D Program of China (Grant No. 2018YFA0306704), the Australian Research Council (Grant No. DP180100691), and the National Science Foundation of China (CN) (Grant No.s 61871111, 12071271).

\bibliography{ref}








\end{document}